\documentclass[final,3p,times,onecolumn]{elsarticle}
\usepackage{slashed}
\usepackage{amsmath,bm}
\allowdisplaybreaks 
\def\bea#1\eea{\begin{align}#1\end{align}} 
\newcommand{\nnu}{\nonumber\\}
\newcommand{\bef}{\begin{figure}[htb]\centering}
\newcommand{\eef}{\end{figure}}
\usepackage{graphicx}
\usepackage{caption}
\usepackage{subcaption}

\usepackage{color}
\usepackage[dvipsnames]{xcolor}
\usepackage[normalem]{ulem}

\newcommand{\beq}{\begin{equation}}
\newcommand{\eeq}{\end{equation}}
\def\bea#1\eea{\begin{align}#1\end{align}}
\newcommand{\be}{\begin{eqnarray}}
\newcommand{\ee}{\end{eqnarray}}

\newcommand{\sla}[1]{{#1}\!\!\!\slash}

\begin{document}

\begin{frontmatter}
\title{Polarized jet fragmentation functions}

\author{Zhong-Bo Kang\fnref{label1,label2,label3}}
\ead{zkang@physics.ucla.edu}
\author{Kyle Lee\fnref{label4,label5}}
\ead{kunsu.lee@stonybrook.edu}
\author{Fanyi Zhao\fnref{label1,label2}}
\ead{fanyizhao@physics.ucla.edu}
\address[label1]{Department of Physics and Astronomy, University of California, Los Angeles, California 90095, USA}
\address[label2]{Mani L. Bhaumik Institute for Theoretical Physics, University of California, Los Angeles, California 90095, USA}
\address[label3]{Center for Frontiers in Nuclear Science, Stony Brook University, Stony Brook, New York 11794, USA}
\address[label4]{C.N. Yang Institute for Theoretical Physics, Stony Brook University, Stony Brook, New York 11794, USA}
\address[label5]{Department of Physics and Astronomy, Stony Brook University, Stony Brook, New York 11794, USA}

\begin{abstract}
We develop the theoretical framework needed to study the distribution of hadrons with general polarization inside jets, with and without transverse momentum measured with respect to the standard jet axis. The key development in this paper, referred to as ``polarized jet fragmentation functions'', opens up new opportunities to study both collinear and transverse momentum dependent (TMD) fragmentation functions. As two examples of the developed framework, we study longitudinally polarized collinear $\Lambda$ and transversely polarized TMD $\Lambda$ production inside jets in both $pp$ and $ep$ collisions. We find that both observables have high potential in constraining spin-dependent fragmentation functions with sizeable asymmetries predicted, in particular, at the future Electron-Ion Collider.
\end{abstract}


\end{frontmatter}

\section{Introduction}
Over the last few years, the study of hadron distributions inside jets has received increasing attention as an effective tool to understand the fragmentation process, describing how the color carrying partons transform into color-neutral particles such as hadrons. Understanding such a fragmentation process is important as it will provide us with a deep insight into the elusive mechanism of hadronization. Theoretical objects which describe the momentum distribution of hadrons inside a fully reconstructed jet is called jet fragmentation functions (JFFs). The usefulness of studying the longitudinal momentum distribution of the hadron in the jet rather than the hadron production itself stems from the former process being differential in the momentum fraction $z_h \equiv p_{hT}/ p_{JT}$, where $p_{hT}$ and $p_{JT}$ are the transverse momenta of the hadron and the jet with respect to the beam axis, respectively. Collinear JFFs in the first process can be matched onto the standard collinear fragmentation functions (FFs), enabling us to extract the usual universal FFs more directly by ``scanning'' the differential $z_h$ dependence. The theoretical developments on the JFFs were first studied in the context of exclusive jet production ~\cite{Procura:2009vm,Jain:2011xz,Jain:2011iu,Chien:2015ctp} and was later extended to the inclusive jet production case ~\cite{Arleo:2013tya,Kaufmann:2015hma,Kang:2016ehg,Dai:2016hzf,Kang:2019ahe}. 

At the same time, the transverse momentum distribution of the hadrons within jets can be sensitive to the transverse momentum dependent fragmentation, described by transverse momentum dependent jet fragmentation functions (TMDJFFs). In ~\cite{Kang:2017glf}, it was demonstrated that such TMDJFFs are closely connected to the standard transverse momentum dependent FFs (TMDFFs)~\cite{Bacchetta:2000jk,Mulders:2000sh,Metz:2016swz} when the transverse momentum of the hadron is measured with respect to the standard jet axis. For the TMD study of the hadron with respect to the Winner-Take-All jet axis, see ~\cite{Neill:2016vbi,Neill:2018wtk}. As for the TMD study inside the groomed jet, see ~\cite{Makris:2017arq,Makris:2018npl,Gutierrez-Reyes:2019msa}. For the recent works on resummation of $\ln z_h$ and $\ln(1-z_h)$, see \cite{Neill:2020bwv,Kaufmann:2019ksh}.

Because of its phenomenological relevance and effectiveness, study of the JFFs has become a very important topic over recent years at the LHC and RHIC, producing measurements for a wide range of identified particles within the jet. Calculations for the JFFs have been performed for single inclusive jet production in unpolarized proton-proton collisions in the context of light charged hadrons ~\cite{Chien:2015ctp,Kaufmann:2015hma,Kang:2016ehg}, heavy-flavor mesons~\cite{Chien:2015ctp,Bain:2016clc,Anderle:2017cgl}, heavy quarkonium~\cite{Kang:2017yde,Bain:2017wvk}, and photons~\cite{Kaufmann:2016nux}. For the relevant experimental results for the LHC and RHIC, see~\cite{Aaboud:2019oac,Aad:2011td,Aad:2011sc,Chatrchyan:2012gw,Chatrchyan:2014ava,Aad:2014wha,Aaboud:2017tke,Aaij:2017fak,Acharya:2019zup,Aaij:2019ctd,Acharya:2018eat,Aad:2019onw,Sirunyan:2019vlp} and~\cite{Adamczyk:2017wld,Adamczyk:2017ynk}, respectively. Study of JFFs is not only important at the LHC and RHIC as already proven to be, but also provides novel insights at the future Electron-Ion Collider~(EIC)~\cite{Accardi:2012qut,Aschenauer:2017jsk,Liu:2018trl,Arratia:2019vju} as we will show below.  

\begin{figure}
    \centering
    \includegraphics[width=2.8in]{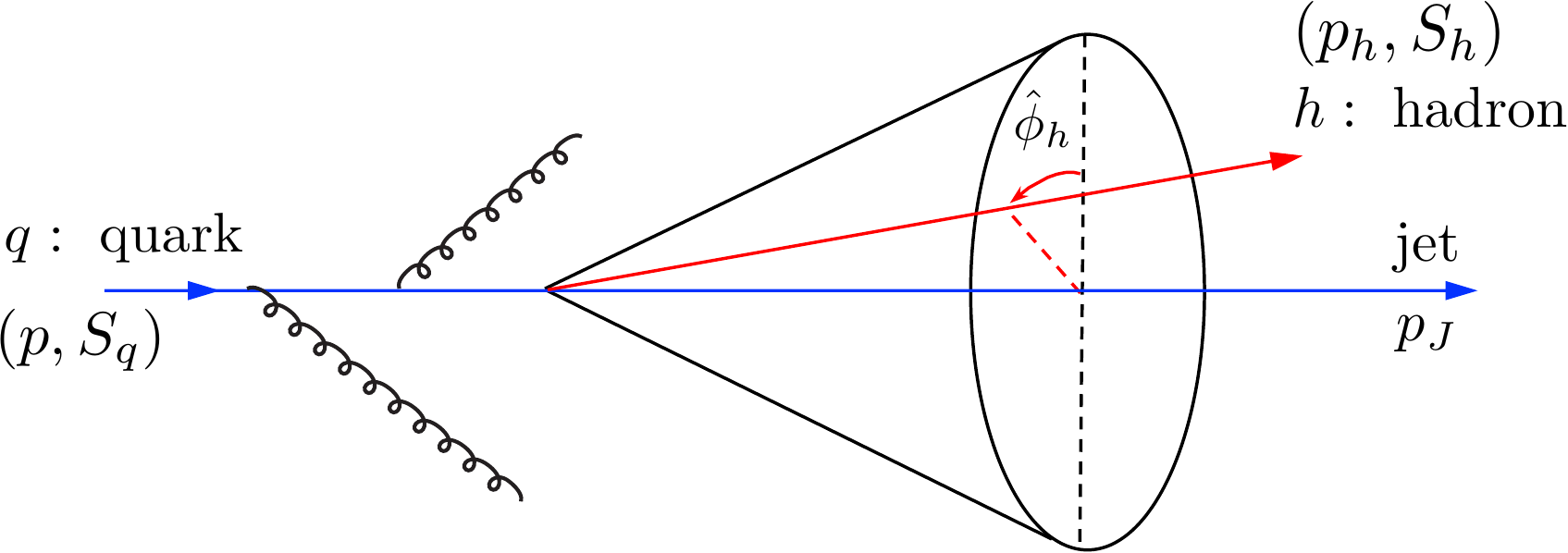}   
    \caption{Illustration for the distribution of hadrons inside a jet, which is initiated by a quark.}
    \label{fig:jet-ff}
\end{figure}

In this paper, we provide the general theoretical framework for studying the distribution of hadrons inside a jet by taking full advantage of the polarization effects. We introduce polarized jet fragmentation functions, where the parton that initiates the jet and the hadron that is inside the jet can both be polarized, as illustrated in fig.~\ref{fig:jet-ff}. We do this in the context of both $pp$ collider like LHC and RHIC, as well as $ep$ collider like the future EIC. Analogous to the standard FFs, we find a slew of different JFFs that have close connection with the corresponding standard FFs. 

When a proton with a general polarization collides with an unpolarized proton or lepton, different JFFs appear with different parton distribution functions (PDFs) and characteristic modulation in the azimuthal angles measured with respect to the scattering plane. Therefore, these observables are not only useful in exploring the spin-dependent FFs, but also in understanding the polarized PDFs. For instance, with the extra handle in $z_h$, we would be able to reduce uncertainties coming from the final state fragmentation functions by restricting to a well-determined $z_h$ region. Alternatively, with well-determined polarized PDFs at hand, we can directly probe spin-dependent FFs through a study of different JFFs. Some applications of spin-dependent JFFs relevant for the RHIC were considered in~\cite{Adamczyk:2017wld,Yuan:2007nd,DAlesio:2010sag,DAlesio:2011kkm,Kang:2017btw,DAlesio:2017bvu}, but other applications are far and wide. To demonstrate this, we consider two phenomenological applications in detail. We demonstrate how one can use spin-dependent JFFs to study the collinear helicity FFs and so-called TMD polarizing fragmentation functions (TMD PFFs). There are, of course, many more possible applications of studying other polarized JFFs which we also list in this paper and will present the details in a forthcoming long paper. Other potential applications include probing the polarization of heavy quarkonium inside the jet~\cite{Kang:2017yde}, which is very promising at the LHC and RHIC. 

The rest of the paper is organized as follows. In section~\ref{sec:theory}, we lay out the theoretical framework for the leading power JFFs with various correlations between the polarization of the hadrons and the fragmenting partons. We provide their physical meaning and connections to the standard fragmentation functions. In section~\ref{sec:application}, we first write down the general structure functions that appear in the cross section with different characteristic modulation in the azimuthal angles. We then use the developed framework to study collinear JFFs describing the longitudinally polarized $\Lambda$-hyperons production at the RHIC and EIC kinematics. We further study transversely polarized $\Lambda$-hyperons production from unpolarized scatterings at the LHC, RHIC, and EIC kinematics. We conclude our paper in section~\ref{sec:conc}.

\section{Theoretical framework}
\label{sec:theory}
In this section, we introduce the concept of polarized jet fragmentation functions. As shown in fig.~\ref{fig:jet-ff}, we consider a fully reconstructed jet which is initiated by a parton, either a quark or a gluon. One further observes a hadron inside the jet, which carries a longitudinal momentum fraction $z_h$ of the jet and a transverse momentum ${\bm j}_\perp$ with respect to the jet axis direction. We consider the general situation where both the parton and the hadron can be polarized. We first provide the definition and parametrization of these novel polarized JFFs, and we then study their connection to the standard FFs.

\subsection{Polarized jet fragmenting functions}
To properly define the momentum and the spin vectors, we define a light-cone vector $n^\mu=(1,0,0,1)$ and its conjugate vector $\bar{n}^\mu=(1,0,0,-1)$, such that $n^2 = \bar{n}^2 = 0$ and $n\cdot\bar{n} = 2$. We can then decompose any four-vector $p^\mu$ as $p^\mu = (p^+, p^-,p_\perp)$ with $p^+ = n\cdot p, p^- = \bar{n}\cdot p$. That is, 
\bea
p^\mu = p^- \frac{n^\mu}{2} + p^+ \frac{\bar{n}^\mu}{2} + p_\perp^\mu\,.
\eea
Let us specify the kinematics of the hadron inside the jet. If the hadron is in a reference frame in which it moves along the $+z$-direction and has no transverse momentum, then the $p_h^-$ component of its momentum would be very large while the $p_h^+$ component is small, $p_h^+\ll p_h^-$. We can parameterize the momentum $p_h$ and the spin vector $S_h$ of the hadron, respectively, as
\bea
p_h = \left(\frac{M_h^2}{p_h^-},p_h^-,0\right)\,,
\qquad
S_h = \left(-\Lambda_h \frac{M_h}{p_h^-},\Lambda_h \frac{p_h^-}{M_h},{\bm S}_{h\perp}\right)\,,
\eea
where $M_h$ is the mass of the hadron, and $\Lambda_h$ and ${\bm S}_{h\perp}$ describe the longitudinal and transverse polarization of the hadron inside the jet, respectively. It is evident that they satisfy the relation $p_h\cdot S_h =0$ as required. 

The general correlators that define jet fragmentation functions in such a hadron frame are given by
\begin{subequations}
\label{eq:correlators}
\bea
\Delta^{h/q}(z, z_h,{\bm j}_\perp, S_h) =& \frac{z}{2N_c} \delta\left(z_h - \frac{p_{h}^-}{p_J^-}\right)
\langle 0| \delta\left(p^- - \bar n\cdot {\mathcal P} \right) \delta^2(\mathcal{P}_\perp/z_h+{\bm j}_\perp)\chi_n(0)  |(Jh)X\rangle \langle (Jh)X|\bar \chi_n(0) |0\rangle,
\\
\Delta^{h/g,\ \mu\nu}(z, z_h, {\bm j}_\perp, S_h) =& \frac{z\,p^-}{(d-2)(N_c^2-1)} \delta\left(z_h - \frac{p_{h}^-}{p_J^-}\right)
\langle 0|  \delta\left(p^- - \bar n\cdot {\mathcal P} \right) \delta^2(\mathcal{P}_\perp/z_h+{\bm j}_\perp){\mathcal B}_{n\perp}^\mu(0) 
 |(Jh)X\rangle  \langle (Jh)X|{\mathcal B}_{n\perp}^\nu(0)  |0\rangle,
\eea
\end{subequations}
for quark and gluon jets, respectively. Here the momenta of the parton that initiates the jet, the jet, and the hadron are given by $p$, $p_J$ and $p_h$, respectively. The energy fractions $z$ and $z_h$ are defined as 
\bea
z = \frac{p_{J}^-}{p^-}, \qquad z_h =  \frac{p_{h}^-}{p_J^-}\,,
\eea
and thus $z$ is the momentum fraction of the parton carried by the jet, while $z_h$ is the momentum fraction of the jet carried by the hadron. Here, we have used the gauge invariant quark and gluon fields, $\chi_n$ and $\mathcal{B}_{n\perp}^\mu$, in the Soft Collinear Effective Theory~\cite{Bauer:2000ew,Bauer:2000yr,Bauer:2001ct,Bauer:2001yt}, with $n$ denoting the direction of the jet in this case. Note that the state $|(Jh)X\rangle$ represents the final-state unobserved particles $X$ and the observed jet $J$ with an identified hadron $h$ inside, denoted collectively by $(Jh)$. Because of this, the equations above also contain the kinematics of the jet, such as the jet radius $R$. We will suppress them here for simplicity, but express them out explicitly when we discuss their evolution equation in the next section. 

Taking into account the spins of both the parton that initiates the jet and the final hadron inside the jet, we can parametrize the above correlators at the leading power as
\begin{subequations}
\label{eq:corr_decomp}
\bea \label{e:qcorr_decomp}
\Delta^{h/q}(z,z_h,{\bm j}_\perp,S_h) & =  
   \Delta^{h/q \, [\slashed{n}]} \, \frac{\slashed{\bar{n}}}{2}
 - \Delta^{h/q \, [\slashed{n} \gamma_5]} \, \frac{\slashed{\bar{n}}\gamma_5}{2} 
 +\Delta^{h/q \, [i n_\nu\,\sigma^{i\nu} \gamma_5]} \, \frac{i\bar{n}_\mu\,\sigma^{i\mu} \gamma_5}{2}
 \,,\\
\label{e:gcorr_decomp}
 \Delta^{h/g,\ ij}(z,z_h,{\bm j}_\perp,S_h) &= \frac{1}{2} \, \delta_T^{ij} \, \Big( \delta_T^{kl} \Delta^{h/g,\ kl} \Big)
 - \frac{i}{2} \, \epsilon_T^{ij} \, \Big( i \epsilon_T^{kl} \Delta^{h/g,\ kl} \Big)
 + \hat{S} \, \Delta^{h/g,\ ij} \,,
\eea
\end{subequations}
where we have $\Delta^{h/q[\Gamma]} \equiv \frac{1}{4} \text{Tr}[\Delta^{h/q}\Gamma]$. The three terms on the r.h.s.~of eq. (\ref{e:qcorr_decomp}) correspond, in order, to unpolarized, longitudinally polarized, and transversely polarized quarks. On the other hand, the three terms on the r.h.s.~of eq.~\eqref{e:gcorr_decomp} correspond to unpolarized, circularly polarized, and linearly polarized gluons. Note that we have used $\delta_T^{ij} = - g_T^{ij}$ with $g_T^{\mu\nu} = g^{\mu\nu} - n^{\mu} \bar{n}^{\nu}/2 - n^{\nu} \bar{n}^{\mu}/2$, $\epsilon_{T}^{\mu\nu} =\epsilon^{\mu\nu\alpha\beta}\bar{n}_\alpha n_\beta /2$, and the symmetrization operator $\hat{S}$ which is defined by~\cite{Meissner:2007rx,Diehl:2001pm, Diehl:2003ny},
\bea
\hat{S}O^{ij}\equiv\frac{1}{2}\left(O^{ij}+O^{ji}-\delta_T^{ij}O^{kk}\right).
\eea

Using the symmetry arguments \cite{Goeke:2005hb,Mulders:1995dh}, the leading TMDJFFs are parametrized for quarks as
\begin{subequations}
\bea
\Delta^{h/q[\sla{n}]}=\ &\mathcal{D}_1^{h/q}(z,z_h,{\bm j}_\perp)-\frac{\epsilon^{ij}_T{ j}_\perp^i{S}_{h\perp}^j}{z_h\,M_h}\mathcal{D}_{1T}^{\perp h/q}(z,z_h,{\bm j}_\perp)\,,\\
\Delta^{h/q[\sla{n}\gamma_5]}=\ &\Lambda_h\, \mathcal{G}_{1L}^{ h/q}(z,z_h,{ j}_\perp)-\frac{{\bm j}_\perp\cdot{\bm S}_{h\perp}}{z_h\,M_h}\mathcal{G}_{1T}^{ h/q}(z,z_h,{\bm j}_\perp)\,,\\
\Delta^{h/q[i n_\nu\,\sigma^{i\nu} \gamma_5]}
=\ &{S}_{h\perp}^i\mathcal{H}_1^{h/q}(z,z_h,{\bm j}_\perp)+\frac{\epsilon^{ij}_T{ j}_\perp^j}{z_h\,M_h}\mathcal{H}_{1}^{\perp h/q}(z,z_h,{\bm j}_\perp)-\frac{{ j}_\perp^i}{z_h\,M_h}\Lambda_h\,{\mathcal{H}_{1L}^{\perp h/q}(z,z_h,{\bm j}_\perp)}\,\nnu
&+\frac{{ j}_\perp^i {\bm j}_\perp\cdot{\bm S}_{h\perp}-\frac{1}{2}{\bm j}_\perp^2S_{h\perp}^i}{z_h^2\,M_h^2}\mathcal{H}_{1T}^{\perp h/q}(z,z_h,{\bm j}_\perp)\,, 
\eea
\end{subequations}
and for gluons as
\begin{subequations}
\label{eq:param}
\bea
\delta_T^{ij} \, \Delta^{h/g,ij}  = & 
\mathcal{D}_1^{h/g}(z,z_h,{\bm j}_\perp)
- \frac{\epsilon_T^{ij} \, {j}_\perp^i \, S_{h\perp}^j}{z_h\,M_h} \, \mathcal{D}_{1T}^{\perp \, h/g}(z,z_h,{\bm j}_\perp) \,,
\\
i \epsilon_T^{ij} \, \Delta^{h/g,ij}  = &
\Lambda_h \, \mathcal{G}_{1L}^{h/g}(z,z_h,{\bm j}_\perp)
- \frac{{\bm j}_\perp\cdot{\bm S}_{h\perp}}{z_h\,M_h} \, \mathcal{G}_{1T}^{h/g}(z,z_h,{\bm j}_\perp)\,,
\\
\hat{S} \, \Delta^{h/g,ij}  
= & \hat{S} \, \bigg[-\frac{{j}_\perp^i \, \epsilon_T^{jk} \, S_{h\perp}^k}{2 z_h\,M_h} \, \mathcal{H}_{1}^{h/g}(z,z_h,{\bm j}_\perp)+ \frac{{j}_\perp^i \, {j}_\perp^j}{2 z_h^2\,M_h^2} \, \mathcal{H}_1^{\perp \, h/g}(z,z_h,{\bm j}_\perp)\nonumber \\
& + \, \frac{{j}_\perp^i \, \epsilon_T^{jk}}{2 z_h^2\,M_h^2} \, \bigg(\Lambda_h\,{j}_\perp^k \, \mathcal{H}_{1L}^{\perp \, h/g}(z,z_h,{\bm j}_\perp)- \left(\frac{{\bm j}_\perp\cdot{\bm S}_{h\perp}{j}_\perp^k-\frac{1}{2}{\bm j}_\perp^2 S_{h\perp}^k}{z_h\,M_h}\right) \, \mathcal{H}_{1T}^{\perp \, h/g}(z,z_h,{\bm j}_\perp)\bigg) \bigg] \,.
\eea
\end{subequations}
Summary of the results with physical interpretation of the quark and gluon TMDJFFs at leading power are given in table~\ref{tabTMDJFF}. Except that they represent the hadron fragmentation inside a fully reconstructed jet, their physical meaning is similar to that of standard TMDFFs as summarized in~\cite{Metz:2016swz}. For this reason, we choose to employ the scripted version of the letters used for the usual TMDFFs  to label TMDJFFs. In this table, U, L, and T represent unpolarized, longitudinally, and transversely polarized states, while Circ and Lin represent circularly and linearly polarized states. Thus $\mathcal{D}_1^{h/q,g}$ describe the distribution of unpolarized hadrons inside a jet which is initiated by an unpolarized quark or gluon. It is also instructive to mention several TMDJFFs that lead to important spin asymmetries in experiments. For example, ${\cal D}_{1T}^{\perp\, h/q, g}$ describes the transverse polarization of hadrons inside the jet that is initiated by an unpolarized quark/gluon, and thus can be used to describe the transverse polarization of Hyperons inside a jet. ${\cal H}_1^{\perp\,h/q}$ is a transversely polarized quark that initiates a jet in which unpolarized hadrons are observed. This function leads to the so-called Collins azimuthal asymmetry for hadrons inside a jet~\cite{Kang:2017btw}. On the other hand, ${\cal H}_1^{\perp\,h/g}$ represents the situation where a  linearly polarized gluon generates a jet in which unpolarized hadrons are observed. 

\begin{table}[t]
    \centering
\begin{tabular}{ |c|c|c|c| } 
 \hline
 h$\backslash$ q & U & L & T \\ 
  \hline
 U & $\mathcal{D}_1^{h/q} $&  &$\mathcal{H}_1^{\perp\ h/q}$  \\ 
  \hline
L &  &  $\mathcal{G}^{h/q}_{1L}$& $\mathcal{H}_{1L}^{\ h/q}$\\ 
  \hline
T & $\mathcal{D}_{1T}^{\perp\ h/q}$ & $\mathcal{G}^{h/q}_{1T}$ & $\mathcal{H}^{h/q}_{1}$, $\mathcal{H}_{1T}^{\perp\ h/q}$\\ 
  \hline
\end{tabular}
\qquad
\begin{tabular}{ |c|c|c|c| } 
 \hline
 h$\backslash$ g & U & Circ & Lin \\ 
  \hline
 U & $\mathcal{D}_1^{h/g} $&  &$\mathcal{H}_1^{\perp\ h/g}$  \\ 
  \hline
L &  &  $\mathcal{G}^{h/g}_{1L}$& $\mathcal{H}_{1L}^{h/g}$\\ 
  \hline
T & $\mathcal{D}_{1T}^{\perp\ h/g}$ & $\mathcal{G}^{h/g}_{1T}$ & $\mathcal{H}^{h/g}_{1}$, $\mathcal{H}_{1T}^{\perp\ h/g}$\\ 
  \hline
\end{tabular}
  \caption{Transverse momentum dependent JFFs for quarks and gluons. Here U, L, and T represent unpolarized, longitudinally, and transversely polarized states, while Circ and Lin represent circularly and linearly polarized states.}
  \label{tabTMDJFF}
\end{table}

Integration over the transverse momentum ${\bm j}_\perp$ dependence formally leads to the collinear JFFs,
\begin{subequations}
\bea  \label{e:D1_int}
\mathcal{D}_1^{h/c}(z,z_h) & =  \int d^2 {{\bm j}}_\perp\, \mathcal{D}_1^{h/c}(z,z_h,{\bm j}_\perp) \,,
\\[0.2cm]  \label{e:G1_int}  
\mathcal{G}_1^{h/c}(z,z_h) & =   \int d^2 {{\bm j}}_\perp \, \mathcal{G}_{1L}^{h/c}(z,z_h,{\bm j}_\perp) \,,
\\  \label{e:H1_int}
\mathcal{H}_1^{h/c}(z,z_h) 
&=  \int d^2 {{\bm j}}_\perp \,  \mathcal{H}_{1}^{h/c}(z,z_h,{\bm j}_\perp)\,,
\eea
\end{subequations}
where $c=q,g$. We assume that the hadron $h$ has a spin-$\frac{1}{2}$, for which there is no integrated linearly polarized gluons as a consequence of conservation of angular momentum, i.e. $\mathcal{H}_1^{h/g}(z,z_h) = 0$. This
is like for PDFs where no collinear gluon transversity exists for a spin-$\frac{1}{2}$ target. Lastly, we note that the correlators in eq.~\eqref{eq:correlators} and parametrizations from eqs.~\eqref{eq:corr_decomp} to \eqref{eq:param} generalize simply to the exclusive jet production case as well, where now the $z$ dependence drops out. See \cite{Procura:2009vm,Jain:2011xz,Jain:2011iu,Chien:2015ctp,Kang:2019ahe} for examples of unpolarized case in exclusive processes. In this paper, we focus on the inclusive jet processes for phenomenological applications.

\subsection{Connection to the standard fragmentation functions}
\label{sub:connection}
Both collinear and transverse momentum dependent JFFs have close relations to the conventional collinear and transverse momentum dependent FFs. For the unpolarized case, the relations between collinear (TMD) JFFs and the standard collinear (TMD) FFs were derived in~\cite{Kang:2016ehg} and~\cite{Kang:2017glf}, respectively. For example, the collinear unpolarized JFFs $\mathcal{D}_1^{h/i}(z,z_h,p_{JT} R,\mu)$ is related to the collinear unpolarized FFs $D_1^{h/i}(z_h,\mu)$ as follows
\bea
\mathcal{D}_1^{h/i}(z,z_h,p_{JT} R,\mu) = \sum_j\int_{z_h}^1 \frac{dz_h'}{z_h'}\mathcal{J}_{ij}(z,z_h',p_{JT} R,\mu)\,D_1^{h/j}\left(\frac{z_h}{z_h'},\mu\right)\,,
\eea
where the coefficient functions $\mathcal{J}_{ij}$ can be found in~\cite{Kang:2016ehg}. Note that we have explicitly included the dependence on the jet kinematics, the jet transverse momentum $p_{JT}$ and the jet radius $R$, as well as the renormalization scale $\mu$. By studying the perturbative behavior of these JFFs, one can derive their renormalization group (RG) equations, which are the same as the usual time-like DGLAP evolution equations, 
\bea
\mu\frac{d}{d\mu}\mathcal{D}_1^{h/i}(z,z_h, p_{JT} R, \mu) = \frac{\alpha_s(\mu)}{\pi} 
 \sum_j \int_z^1  \frac{dz'}{z'}P_{ji}\left(\frac{z}{z'}, \mu \right) \mathcal{D}_1^{h/j}(z',z_h, p_{JT} R, \mu)\,,
\eea
where $P_{ji}$ are the splitting functions for unpolarized fragmentation functions~\cite{Altarelli:1977zs,Stratmann:1996hn}. We will derive all the other relations between TMDJFFs and TMDFFs, as well as their corresponding collinear version in the future publication~\cite{KangLeeZhao:inprogress}. Here we only list the coefficients that are relevant to the key spin observables in the next section.

In order to study longitudinal polarization of a $\Lambda$ particle inside a jet produced in single longitudinally polarized proton-proton and/or proton-lepton collisions, $\vec{p}+\left(p/e\right)\to (\text{jet}\vec{\Lambda})+X$, one would need the polarized collinear JFFs $\mathcal{G}_1^{h/i}(z,z_h, p_{JT} R, \mu)$. The RG equations are given by
\bea
\mu\frac{d}{d\mu}\mathcal{G}_1^{h/i}(z,z_h, p_{JT} R, \mu) = \frac{\alpha_s(\mu)}{\pi} 
 \sum_j \int_z^1  \frac{dz'}{z'} \Delta_LP_{ji}\left(\frac{z}{z'}, \mu \right) \mathcal{G}_1^{h/j}(z',z_h, p_{JT} R, \mu)\,,
\eea
where $\Delta_L P_{ji}(z_h)$ are the time-like splitting functions for helicity fragmentation functions~\cite{Altarelli:1977zs,Stratmann:1996hn}. This is expected since $\mathcal{G}_1^{h/i}$ describes longitudinally polarized $\Lambda$ production inside a jet, which is initiated by a longitudinally polarized parton. Likewise, $\mathcal{G}_1^{h/i}$ can be matched onto helicity FFs $G_1^{h/i}(z_h, \mu)$ as follows 
\bea
\label{eq:longmatch}
\mathcal{G}_1^{h/i}(z,z_h,p_{JT} R,\mu) = \sum_j\int_{z_h}^1 \frac{dz_h'}{z_h'}\mathcal{J}^L_{ij}(z,z_h',p_{JT} R,\mu)\,G_1^{h/j}\left(\frac{z_h}{z_h'},\mu\right)\,,
\eea
where the matching coefficients $\mathcal{J}^L_{ij}$ at the next-to-leading order (NLO) for jets reconstructed via the anti-$k_T$ algorithm are given by 
\begin{subequations}
\bea
\mathcal{J}^L_{qq}(z,z_h,p_{JT} R,\mu)&=\delta(1-z)\delta(1-z_h)+\frac{\alpha_s}{2\pi}\left\{\ln\left(\frac{\mu^2}{p_{JT}^2R^2}\right)\left[\Delta_L P_{qq}(z)\delta(1-z_h)-\Delta_L P_{qq}(z_h)\delta(1-z)\right]\right.\nonumber\\
&+\delta(1-z)\left[2C_F(1+z_h^2)\left(\frac{\ln{(1-z_h)}}{1-z_h}\right)_++C_F(1-z_h)+\Delta_L\mathcal{I}_{qq}^{\text{anti}-k_T}(z_h) \right]\nonumber\\
&\left.-\delta(1-z_h)\left[2C_F(1+z^2)\left(\frac{\ln{(1-z)}}{1-z}\right)_++C_F(1-z)\right]\right\} \,, \\
\mathcal{J}^L_{qg}(z,z_h,p_{JT} R,\mu)&=\frac{\alpha_s}{2\pi}\left\{\ln\left(\frac{\mu^2}{p_{JT}^2R^2}\right)\bigg[\Delta_L P_{gq}(z)\delta(1-z_h)-\Delta_L P_{gq}(z_h)\delta(1-z)\bigg]\right.\nonumber\\
&+\delta(1-z)\left[2\Delta_L P_{gq}(z_h)\ln (1-z_h)-2C_F(1-z_h)+\Delta_L\mathcal{I}_{gq}^{\text{anti}-k_T}(z_h) \right]\nonumber\\
&-\delta(1-z_h)\bigg[2\Delta_L P_{gq}(z)\ln (1-z)-2C_F(1-z)\bigg]\Bigg\} \,, \\
\mathcal{J}^L_{gq}(z,z_h,p_{JT} R,\mu)&=\frac{\alpha_s}{2\pi}\left\{\ln\left(\frac{\mu^2}{p_{JT}^2R^2}\right)\left[\Delta_L P_{qg}(z)\delta(1-z_h)-\Delta_L P_{qg}(z_h)\delta(1-z)\right]\right.\nonumber\\
&+\delta(1-z)\left[2\Delta_L P_{qg}(z_h)\ln(1-z_h)+2T_F(1-z_h)+\Delta_L\mathcal{I}_{qg}^{\text{anti}-k_T}(z_h) \right]\nonumber\\
&-\delta(1-z_h)\bigg[2\Delta_L P_{qg}(z)\ln (1-z)+2T_F(1-z)\bigg]\Bigg\} \,, \\
\mathcal{J}^L_{gg}(z,z_h,p_{JT} R,\mu)&=\delta(1-z)\delta(1-z_h)+\frac{\alpha_s}{2\pi}\left\{\ln\left(\frac{\mu^2}{p_{JT}^2R^2}\right)\left[\Delta_L P_{gg}(z)\delta(1-z_h)-\Delta_L P_{gg}(z_h)\delta(1-z)\right]\right.\nonumber\\
&+\delta(1-z)\left[4N_c\left(2(1-z_h)^2+z_h\right)\left(\frac{\ln (1-z_h)}{1-z_h}\right)_+-4N_c(1-z_h)+\Delta_L\mathcal{I}_{gg}^{\text{anti}-k_T}(z_h)\right]\nonumber\\
&\left.-\delta(1-z_h)\left[4N_c\left(2(1-z)^2+z\right)\left(\frac{\ln (1-z)}{1-z}\right)_+-4N_c(1-z)\right]\right\} \,,
\eea
\end{subequations}
with the anti-$k_T$ algorithm term
\bea
\Delta_L\mathcal{I}_{ji}^{\text{anti}-k_T}(z_h)&=2\Delta_L P_{ji}(z_h)\ln z_h\,.
\eea

Now if one wants to explore the transverse momentum ${\bm j}_\perp$ distribution of hadrons inside the jet, along with any spin-dependent correlations, one would need the TMDJFFs. For example, if one wants to measure both the unpolarized and transversely polarized $\Lambda$ production inside the jet in unpolarized proton-proton and/or proton-lepton collisions, $p+\left(p/e\right)\to (\text{jet}{\Lambda^\uparrow})+X$, one would need TMDJFFs $\mathcal{D}_{1}^{h/i}(z, z_h, {\bm j}_\perp, p_{JT}R, \mu)$ and $\mathcal{D}_{1T}^{\perp\, h/i}(z, z_h, {\bm j}_\perp, p_{JT}R, \mu)$, respectively. It has been demonstrated in~\cite{Kang:2017glf} that unpolarized TMDJFF $\mathcal{D}_{1}^{h/i}$ is related to the standard TMDFF $D_{1}^{h/i}$. Following the same procedure, we show only the final results here and leave the details in the future publication. First of all, since $\mathcal{D}_{1T}^{\perp h/i}$ involves an unpolarized parton that initiates the jet, its evolution follows the same RG equations as the unploarized case
\bea
\mu\frac{d}{d\mu}\mathcal{D}_{1T}^{\perp\,h/i}(z,z_h,{\bm j}_\perp, p_{JT} R, \mu) = \frac{\alpha_s(\mu)}{\pi} 
 \sum_j \int_z^1  \frac{dz'}{z'} P_{ji}\left(\frac{z}{z'}, \mu \right) \mathcal{D}_{1T}^{\perp\,h/j}(z',z_h, {\bm j}_\perp, p_{JT} R, \mu)\,,
\eea
which allow us to evolve $\mathcal{D}_{1T}^{\perp\,h/i}$ from the typical jet scale $\mu_J\sim p_{JT} R$ to the hard scale $\mu\sim p_{JT}$ and thus resums the logarithm of the jet radius $R$. On the other hand, at the scale $\mu_J$, we have  
\bea
\mathcal{D}_{T}^{\perp\, h/i}(z,z_h,{\bm j}_\perp,p_{JT} R,\mu_J) = \mathcal{C}_{i\to j}(z,p_{JT} R,\mu_J)\,D_{1T}^{\perp\, h/j}(z_h,{\bm j}_\perp;\mu_J)\,,
\label{eq:PFFs}
\eea
where $D_{1T}^{\perp\, h/j}$ is the TMD PFFs that describes fragmentation of an unpolarized parton into a transversely polarized hadron. On the other hand, the coefficient functions $\mathcal{C}_{i\to j}$ are the same as those in the unpolarized case and can be found in~\cite{Kang:2017glf}. 

\section{Application of polarized jet fragmentation functions}
\label{sec:application}
In this section, we demonstrate the application of the polarized jet fragmentation functions defined above and show how they can bring novel insights into the study of spin and TMD effects. We first set up the general framework, and we then choose two specific observables to show the numerical signatures at RHIC, LHC, and the future EIC.

\subsection{General structure of the observables}
\label{sub:general}
\begin{figure}[htb]
    \centering
    \includegraphics[width=4in]{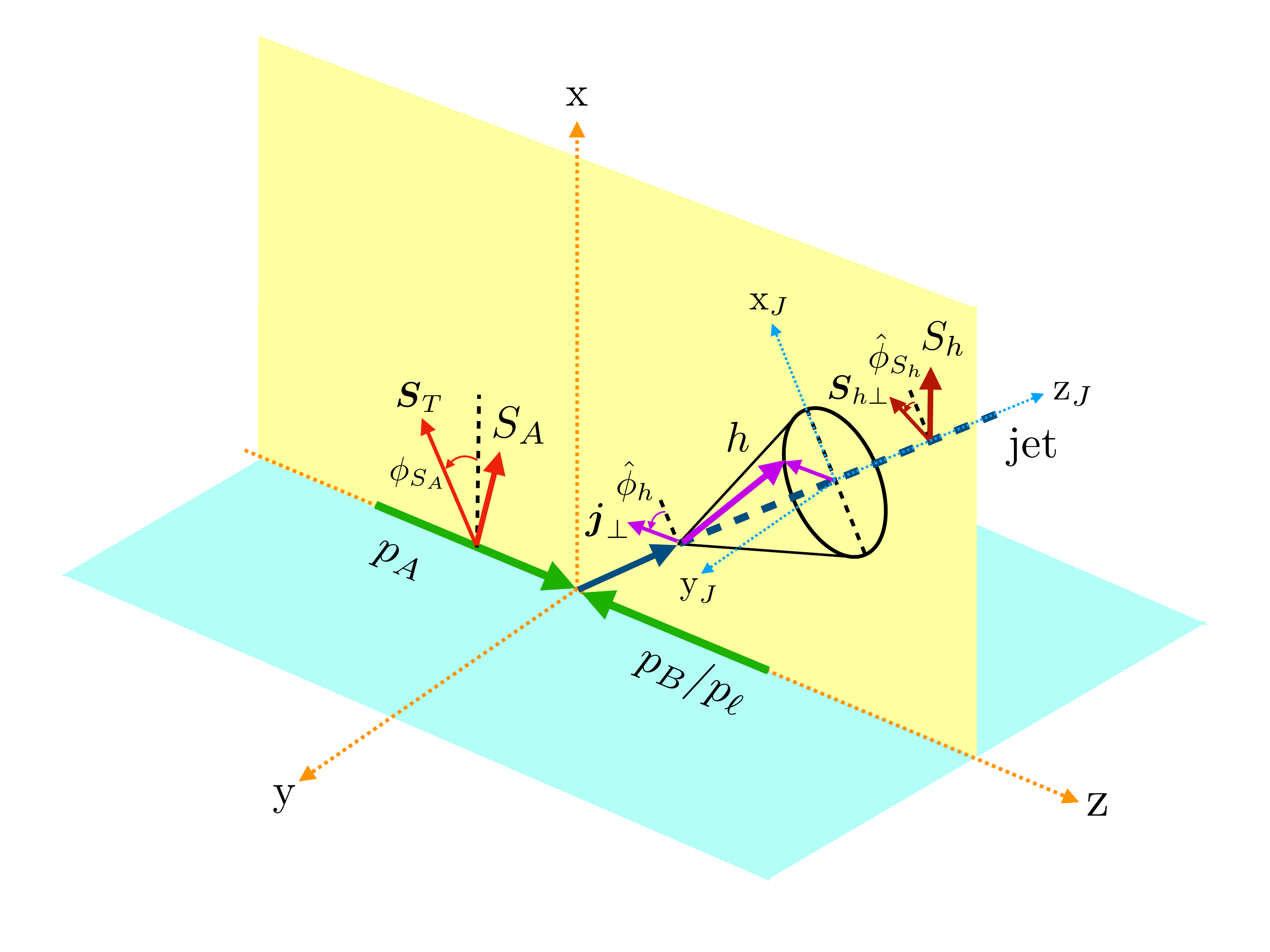}   
    \caption{Illustration for the distribution of hadrons inside jets in the collisions of a polarized proton and an unpolarized proton or lepton.}
    \label{fig:illu}
\end{figure}

We consider hadron distribution inside jets in either lepton-proton or proton-proton collisions, as illustrated in fig.~\ref{fig:illu},
\bea
p(p_A,{S}_{A}) + \Big(p(p_B) / e(p_\ell)\Big) \rightarrow \left(\text{jet}(\eta_J, p_{JT}, R)\ h(z_h,{\bm j}_\perp, S_h)\right)+X\,,
\eea
where a polarized proton with the spin $S_A$ and momentum $p_A$ (moving along ``$+z$'' direction) scatters on an unpolarized proton (or lepton) with momentum $p_B$ ($p_\ell$) (moving along ``$-z$'' direction) and produces a jet reconstructed in the usual anti-$k_T$ algorithm~\cite{Cacciari:2008gp} with the jet radius parameter $R$, rapidity $\eta_J$, and transverse momentum $p_{JT}$. One further observes a hadron inside the jet with its spin $S_h$, which carries a longitudinal momentum fraction $z_h$ of the jet and a transverse momentum ${\bm j}_\perp$ with respect to the standard jet axis direction. The spin vector of the incoming proton in the center-of-mass frame is parametrized as
\bea
S_A &= \left(-\lambda \frac{M_A}{p_A^-},\lambda \frac{p_A^-}{M_A},{\bm S}_T\right)\,,
\eea
where $\lambda$ is the helicity of the polarized proton with mass $M_A$, while ${\bm S}_T$ is its transverse spin vector. Needless to say, the observables in proton-proton collisions can be studied in both RHIC and LHC facilities, while we have EIC in mind for those in lepton-proton collisions. 

In this paper, we consider the case where the jet fragmentation function measurement is performed for an inclusive jet sample in proton-proton (lepton) collisions, where the sum over all particles in the final state $X$ besides the observed jet is performed. In this case, it is the so-called semi-inclusive JFFs defined in the previous section that are probed. On the other hand, if one performs the measurements in an exclusive jet production, such as the back-to-back lepton-jet in lepton-proton collisions, or $Z$-jet correlations in proton-proton collisions, one should instead use the exclusive version of the JFFs. For details, see e.g.~\cite{Kang:2019ahe,Liu:2018trl,Buffing:2018ggv}.

For the scattering illustrated in fig.~\ref{fig:illu}, and for single inclusive jet production, the most general azimuthal dependence for the hadron distribution inside the jet differential in both $z_h$ and ${\bm j}_\perp$ can be written in the following form
\bea
\label{eq:gen_str}
\frac{d\sigma^{p(S_A)+p/e\to (\text{jet}\,h(S_h))X}}{dp_{JT} d\eta_{J} dz_h d^2{\bm j}_\perp}=&F_{UU,U}+|{\bm S}_T|\sin(\phi_{S_A}-\hat{\phi}_h)F^{\sin(\phi_{S_A}-\hat{\phi}_h)}_{TU,U}+\Lambda_h\left[\lambda F_{LU,L}+|{\bm S}_T|\cos(\phi_{S_A}-\hat{\phi}_h)F^{\cos(\phi_{S_A}-\hat{\phi}_h)}_{TU,L}\right]\nnu
&+|{\bm S}_{h\perp}|\bigg\{\sin(\hat{\phi}_h-\hat{\phi}_{S_h})F^{\sin(\hat{\phi}_h-\hat{\phi}_{S_h})}_{UU,T}+\lambda\cos(\hat{\phi}_h-\hat{\phi}_{S_h}) F^{\cos(\hat{\phi}_h-\hat{\phi}_{S_h})}_{LU,T}\nnu
&\qquad\quad+|{\bm S}_T|\big(\cos(\phi_{S_A}-\hat{\phi}_{S_h})F^{\cos(\phi_{S_A}-\hat{\phi}_{S_h})}_{TU,T}+\cos(2\hat{\phi}_h-\hat{\phi}_{S_h}-\phi_{S_A})F^{\cos(2\hat{\phi}_h-\hat{\phi}_{S_h}-\phi_{S_A})}_{TU,T}\big)\bigg\}\,,
\eea
where $F_{AB,C}$ are the spin-dependent structure functions, with $A, B$, and $C$ indicating the polarization of the polarized incoming proton, the unpolarized incoming proton or lepton, and the outgoing hadron inside the jet, respectively. Note that the situation where both incoming protons (or proton-lepton) are polarized have been extensively studied in the literature~\cite{Aidala:2012mv}, and can be used to explore the polarized PDFs. Our focus here is to understand how the polarization in the fragmentation functions inside the jet can couple with the polarization in the incoming proton and thus to figure out the novel insights that come with these observables. Explicitly, we note that
\bea
F_{UU,C} \sim f_{a/A}(x_a, \mu)\,,
\qquad
F_{LU,C} \sim g_{a/A}(x_a, \mu)\,,
\qquad
F_{TU,C} \sim h_{a/A}(x_a, \mu)\,,
\eea
where $f_{a/A},g_{a/A}$, and $h_{a/A}$ stand for unpolarized, helicity, and transversity collinear PDFs in the incoming proton, respectively. As the transverse momentum is measured with respect to the jet axis, its dependence on the inherent transverse momentum of the partons from the incoming hadrons is power suppressed~\cite{Kang:2016ehg}. Therefore, incoming distributions remain collinear whether we have the transverse momentum measured or not for the final hadron inside the jet. Then each hadron in $A= U,L,T$ state create unpolarized, longitudinally (circularly), and transversely (linearly) polarized quarks (or gluons) respectively that fragments into hadron of state $C$. Comparing this fact with the table~\ref{tabTMDJFF} show which JFFs each $F_{AB,C}$ are sensitive to. That is,
\begin{subequations}
\bea
&F_{UU,U} \sim f_{a/A} \,\mathcal{D}_1^{h/c}\,,
\
F^{\sin(\hat{\phi}_h-\hat{\phi}_{S_h})}_{UU,T} \sim f_{a/A}\,\mathcal{D}_{1T}^{\perp h/c}\,,
\\
&F_{LU,L} \sim g_{a/A}\,\mathcal{G}_1^{h/c}\,,
\
F^{\cos(\hat{\phi}_h-\hat{\phi}_{S_h})}_{LU,T} \sim g_{a/A}\,\mathcal{G}_{1T}^{h/c}\,,
\\
&F^{\sin(\phi_{S_A}-\hat{\phi}_h)}_{TU,U} \sim h_{a/A}\,\mathcal{H}^{\perp h/c}\,,
\
F^{\cos(\phi_{S_A}-\hat{\phi}_h)}_{TU,L} \sim h_{a/A}\,\mathcal{H}_{1L}^{h/c}\,,
\
F^{\cos(\phi_{S_A}-\hat{\phi}_{S_h})}_{TU,T} \sim h_{a/A}\,\mathcal{H}_1^{h/c}\,,
\
F^{\cos(2\hat{\phi}_h-\hat{\phi}_{S_h}-\phi_{S_A})}_{TU,T}  \sim h_{a/A}\,\mathcal{H}_{1T}^{\perp h/c}\,,
\eea
\end{subequations}
where we have suppressed the collinear unpolarized PDFs $f_{b/B}$ in the unpolarized incoming proton. Detailed expressions of different structure functions will be given in the forthcoming long paper~\cite{KangLeeZhao:inprogress}. 

It is instructive to point out that the azimuthal angle $\phi_{S_A}$ of the transverse spin ${\bm S}_T$ of the incoming proton is measured in the center-of-mass (CM) frame. As shown in fig.~\ref{fig:illu}, in the CM frame, we align the jet momentum in the $x$-$z$ plane. At the same time, we choose another reference frame in which the jet momentum aligns in $+z_J$ direction, with the $x_J$ axis in the $x$-$z$ plane. We then measure the azimuthal angle of the relative momentum ${\bm j}_\perp$ and the transverse spin vector ${\bm S}_{h\perp}$ of the hadron in the jet in this $x_J$-$y_J$-$z_J$ reference frame, and they are denoted as $\hat{\phi}_h$ and $\hat{\phi}_{S_h}$, respectively. 

If one integrates over ${\bm j}_\perp$ and thus only measures the $z_h$-distribution of hadrons inside the jet, one would have
\bea
\label{eq:long_str}
\frac{d\sigma^{p(S_A)+p/e\to (\text{jet}\,h(S_h))X}}{dp_{JT} d\eta_{J} dz_h}=&F_{UU,U}+\Lambda_h\,\lambda F_{LU,L}+|{\bm S}_{h\perp}||{\bm S}_T|\cos(\phi_{S_A}-\hat{\phi}_{S_h})F^{\cos(\phi_{S_A}-\hat{\phi}_{S_h})}_{TU,T}\,,
\eea
where $F_{AB,C}$ are the collinear version of the structure functions given in eq.~\eqref{eq:gen_str}. Of course if we choose to polarize both incoming particles, there would be more terms in both eqs.~\eqref{eq:gen_str} and~\eqref{eq:long_str}. In the remaining section, as key examples of application, we study longitudinal spin transfer as encoded in the ${\bm j}_\perp$-integrated version of $F_{LU,L}$, as well as the transverse polarization of $\Lambda$ production as contained in the structure function $F^{\sin(\hat{\phi}_h-\hat{\phi}_{S_h})}_{UU,T}$. The unpolarized hadron distribution in jets as encoded in $F_{UU,U}$~\cite{Kang:2016ehg,Kang:2017glf} and the hadron Collins asymmetry as encoded in $F^{\sin(\phi_{S_A}-\hat{\phi}_{h})}_{TU,U}$~\cite{Kang:2017btw} have been explored previously.

\subsection{Example 1: Longitudinally polarized $\Lambda$}
\label{sec:helicity}
As their polarizations can be determined via dominant weak decay channel  $\Lambda \to p\,\pi\,(\bar{\Lambda} \to \bar{p}\,\pi)$, $\Lambda (\bar{\Lambda})$-hyperons are particularly suited for studying the spin-dependent fragmentation. There have been many different measurements of polarized $\Lambda (\bar{\Lambda})$-hyperons, and we give predictions for longitudinally polarized $\Lambda$ inside jets in this section.

\begin{figure}
    \centering
    \includegraphics[width=6in]{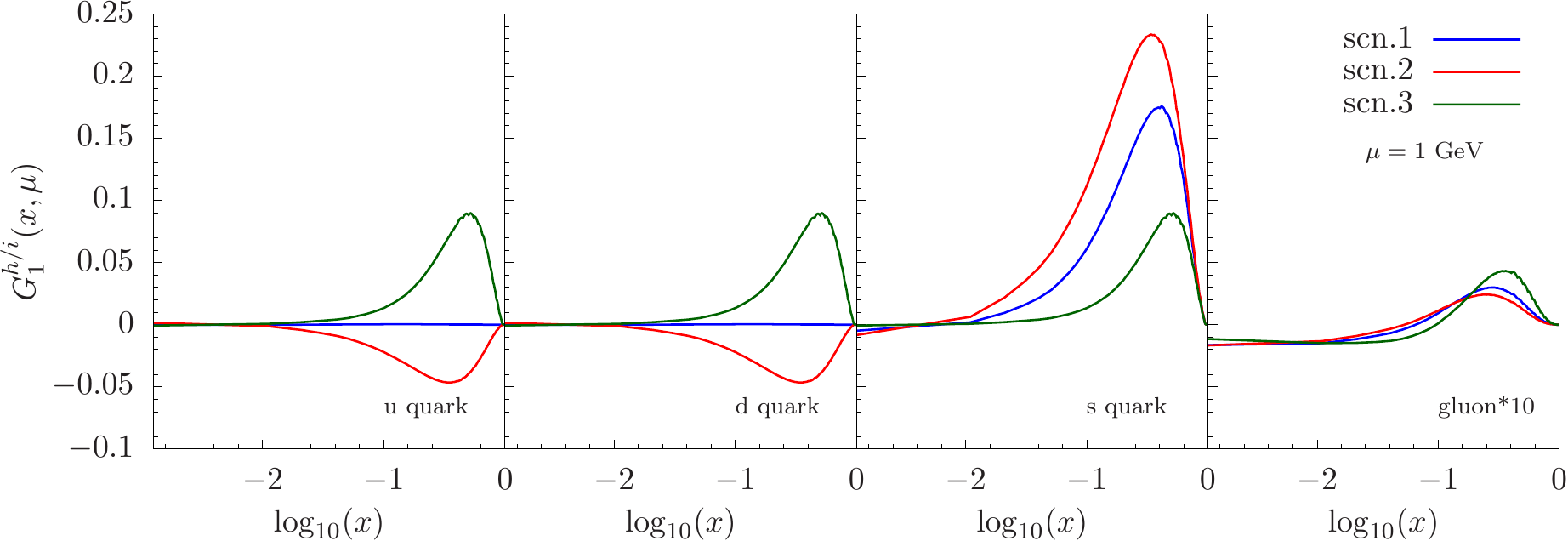}   
    \caption{The longitudinally polarized fragmentation functions plotted at $\mu = 1~\text{GeV}$ for various scenarios \cite{deFlorian:1997zj} consistent with the LEP data. Scenario 1 has only polarized $s$ quark nonvanishing, scenario 2 has $u$ and $d$ quark equal, but with opposite sign from $s$ quark, scenario 3 has $u,~d$, and $s$ quark equal to each other. In all scenarios, gluon vanishes at the input scale and is generated entirely from QCD evolution.}
    \label{fig:polFF}
\end{figure}

Longitudinal spin dependent $\Lambda/\bar{\Lambda}$ fragmentation functions were first determined (see also~\cite{Burkardt:1993zh} for earlier study) by analyzing the LEP data~\cite{Buskulic:1996vb} to NLO \cite{deFlorian:1997zj}, which was not able to constrain the valence fragmentation functions for all flavors; vastly different helicity FFs motivated by different scenarios in fig.~\ref{fig:polFF} were able to describe the LEP data equally well. As demonstrated by~\cite{deFlorian:1998ba}, however, study of the longitudinal spin transfer $D_{LL}$ in longitudinally polarized proton-proton collisions, $\vec{p} p\to \vec{\Lambda}X$, as a function of rapidity can discriminate the different scenarios of valence spin dependent fragmentation functions. The positive rapidity region (corresponding to the forward region of the polarized proton) correspond to the valence region of the polarized proton, and thus spin transfers are dominated by $u$ and $d$ quarks inside the polarized proton. Since these polarized $u$ and $d$ quarks then fragment, $D_{LL}$ plotted as a function of rapidity can distinguish different scenarios of helicity FFs of $u$ and $d$ quark. For the past decade, there have been many measurements and analyses done for such longitudinal spin transfer to $\Lambda$ and $\bar{\Lambda}$ hyperons by the STAR collaboration at RHIC~\cite{Xu:2006my,Abelev:2009xg,Adam:2018kzl}. However, the measurements are only binned in two rapidity bins, the negative ($-1.2 < \eta_{J} < 0$) and positive ($0 < \eta_{J} < 1.2$) bins (at the most recent measurement \cite{Adam:2018kzl}), and is not yet able to discriminate the different scenarios proposed in~\cite{deFlorian:1997zj}. 

As the measurement can be binned more continuously in the transverse momentum $p_{\Lambda T}$ of the $\Lambda$ particle, we propose using the relevant collinear helicity JFF to scan through the helicity FFs more directly by studying $z_\Lambda = p_{\Lambda T}/p_{J T}$ distribution of the longitudinally polarized $\Lambda$ particles inside a jet in longitudinally polarized proton-proton collisions, $\vec{p}+p\to (\text{jet}\vec{\Lambda})+X$. Analogous longitudinal spin transfer $D_{LL}^{\text{jet}\Lambda}$ for $\Lambda$ polarization in the jet can be defined as follows
\bea
\label{eq:assyjet}
D_{LL}^{\text{jet}\Lambda} = \frac{d\Delta\sigma^{\vec{p}p\to  (\mathrm{jet}\vec{\Lambda})X}}{dp_{JT}d\eta_J dz_\Lambda}   \Bigg/ \frac{d\sigma^{pp\to  (\mathrm{jet}\Lambda)X}}{dp_{JT}d\eta_J dz_\Lambda} = \frac{F_{LU,L}}{F_{UU,U}}  \,,
\eea
where the last equation uses the structure functions in eq.~\eqref{eq:long_str}. For the unpolarized cross section in the denominator, factorization is given by~\cite{Kang:2016ehg}~\footnote{we will write factorization expressions for $pp$, but similar expressions for $ep$ follow simply.}
\bea
\frac{d\sigma^{pp\to  (\mathrm{jet}\Lambda)X}}{dp_{JT}d\eta_{J} dz_\Lambda} =\sum_{a,b,c}\int_{x_a^{\text{min}}}^1\frac{dx_a}{x_a}f_a(x_a,\mu)\int_{x_b^{\text{min}}}^1\frac{dx_b}{x_b}f_b(x_b,\mu)\int_{z_c^{\text{min}}}^1\frac{dz_c}{z_c^2}H_{ab}^c \,\mathcal{D}_{1}^{h/c}(z_c,z_\Lambda,p_{JT} R,\mu)\,.
\eea
On the other hand, the numerator of eq.~\eqref{eq:assyjet} is defined as $d\Delta\sigma = [d\sigma(+,+) - d\sigma(+,-)]/2$ with the first and second index representing the helicities $\lambda$ and $\Lambda_\Lambda$, respectively, and it can be written in the following factorized form
\bea
\label{eq:longfact}
\frac{d\Delta\sigma^{\vec{p}p\to  (\mathrm{jet}\vec{h})X}}{dp_{JT}d\eta_J dz_\Lambda}= \sum_{a,b,c}\int_{x_a^{\text{min}}}^1\frac{dx_a}{x_a}g_a(x_a,\mu)\int_{x_b^{\text{min}}}^1\frac{dx_b}{x_b}f_b(x_b,\mu)\int_{z_c^{\text{min}}}^1\frac{dz_c}{z_c^2}\,\Delta_{LL}H_{ab}^{c} \,\mathcal{G}_1^{h/c}(z_c,z_\Lambda,p_{JT}R,\mu)\,.
\eea
Here $f_{a}(x_a, \mu)$ and $g_a(x_a, \mu)$ are the unpolarized PDFs and helicity parton distribution functions, and $H_{ab}^c$ (or $\Delta_{LL} H_{ab}^c$) are the corresponding hard functions, respectively. Finally, $\mathcal{D}_1^{h/c}$ and $\mathcal{G}_1^{h/c}$ are the relevant unpolarized collinear JFFs and helicity JFFs defined in the previous section. Once again, $\mathcal{G}_1^{h/c}$ can be matched onto the standard helicity FFs as in eq.~\eqref{eq:longmatch}.

Let us now present the numerical results for longitudinally polarized $\Lambda$ production inside jets at the RHIC and EIC kinematics. We use the polarized and unpolarized NNPDF sets~\cite{ball2013unbiased,ball2013parton} for collinear helicity parton distributions and unpolarized PDFs, respectively. For both unpolarized and longitudinally polarized fragmentation functions for $\Lambda$, we use the leading-order set from \cite{deFlorian:1997zj}, whose three different scenarios are plotted in fig.~\ref{fig:polFF}. 

\begin{figure}[htb]
\centering
\begin{subfigure}{.5\textwidth}
  \centering
  \includegraphics[width=.85\linewidth]{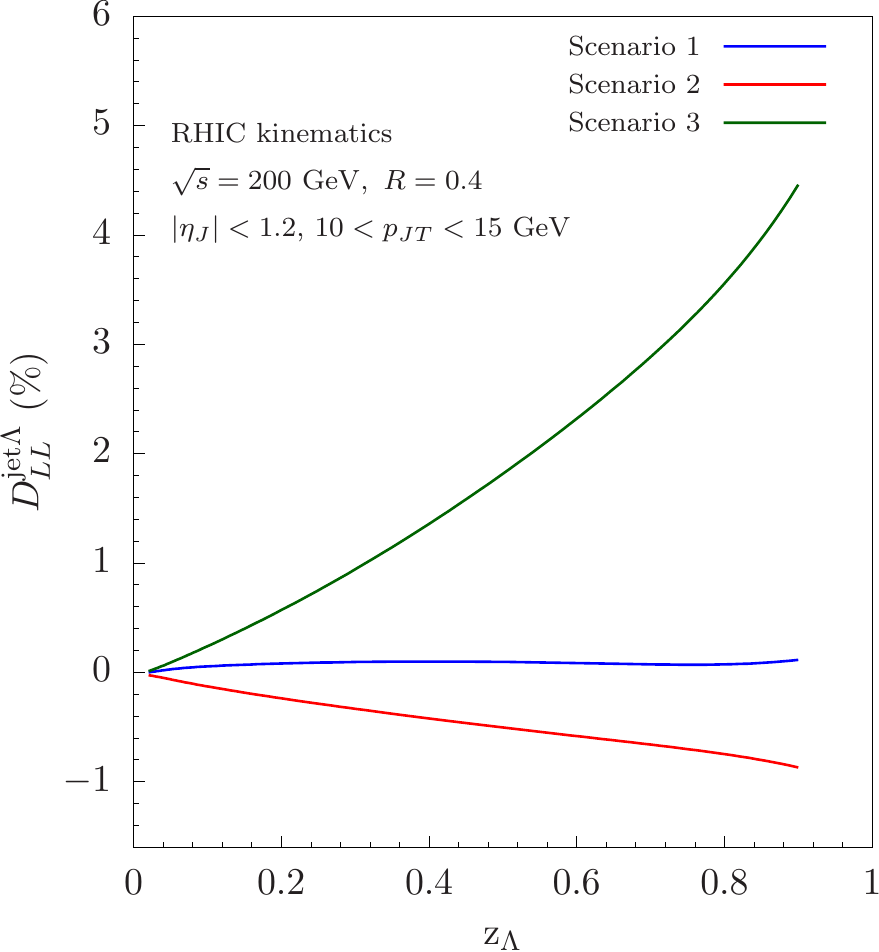} 
  \caption{}
    \label{fig:DLLpp}
\end{subfigure}%
\begin{subfigure}{.5\textwidth}
  \centering
  \includegraphics[width=.85\linewidth]{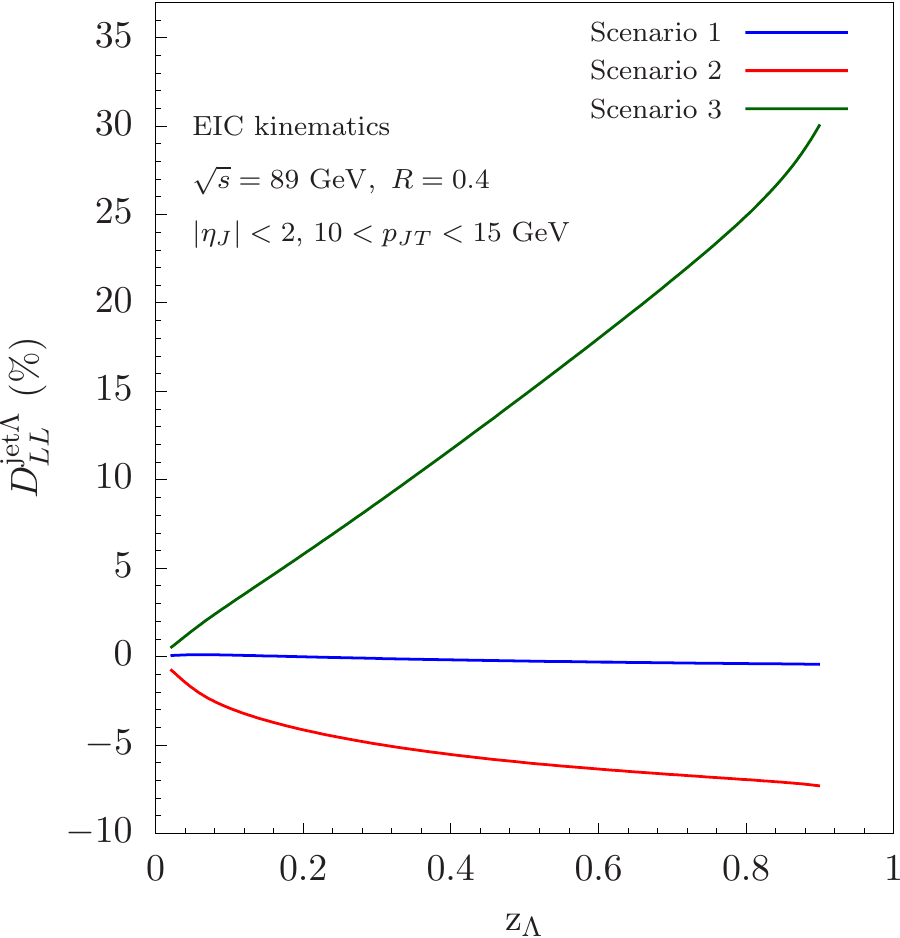} 
  \caption{}
  \label{fig:DLLep}
\end{subfigure}
\caption{Predictions for the asymmetry $D_{LL}^{\text{jet}\Lambda}$ at the RHIC (left) and EIC (right) for various scenarios, which have been proposed in~\cite{deFlorian:1997zj} and shown in fig.~\ref{fig:polFF}. Our predictions show that $D_{LL}^{\text{jet}\Lambda}$ is capable of discriminating different scenarios.}
\end{figure}

Following the recent STAR measurement at RHIC~\cite{Adam:2018kzl}, we first study $\Lambda$ polarization inside jets that are reclustered through the anti-$k_T$ algorithm with a jet radius $R=0.4$ in proton-proton collisions at $\sqrt{s} = 200$ GeV. We plot our predictions for $D_{LL}^{\text{jet}\Lambda}$ in fig.~\ref{fig:DLLpp} as a function of $z_\Lambda$, while integrating over the jet transverse momentum and rapidity: $10~<p_{JT} <~15$ GeV and $|\eta_J| < 1.2$. We find that the size of the asymmetry is on the order of a few percents, and polarized FFs in different scenarios lead to quite different behavior in the spin asymmetry. Such behavior can be understood from the fact that contributions from $u$ and $d$ quarks dominate in this kinematic region, and thus $D_{LL}^{\text{jet}\Lambda}$ follows the signs of the corresponding helicity FFs $G_1^{\Lambda/u,d}$. In other words, as $z_\Lambda$ directly probes the polarized FFs in fig.~\ref{fig:polFF}, no other differential information is required to discriminate the three scenarios. This new observable is thus capable of discriminating different scenarios without binning in $\eta_J$, as in the case of single inclusive $\Lambda$ production~\cite{deFlorian:1998ba}. 

Next, we study longitudinally polarized $\Lambda$ for the EIC at a CM energy of $\sqrt{s}=89~$GeV. We measure $R=0.4$ anti-$k_T$ jets with transverse momentum and rapidity in the range $10~<p_{JT} <~15$ GeV and $|\eta_J| < 2$, respectively. We notice that with quark contributions enhanced by the LO $eq \to eq$ process, $D_{LL}^{\text{jet}\Lambda}$ becomes order of tens of percents at the EIC as shown in fig.~\ref{fig:DLLep} and will be able to clearly discriminate different scenarios of helicity FFs. 

\subsection{Example 2: Transversely polarized $\Lambda$ from unpolarized scatterings}
\label{sec:PFFs}
As a phenomenological application of polarized TMDJFFs, we consider production of a transversely polarized $\Lambda$ inside a jet from an unpolarized fragmenting parton. The corresponding standard TMDFFs are the TMD PFFs $D_{1T}^{\perp\,h/c}$, as metnioned in section~\ref{sub:connection}. The Belle Collaboration has recently measured transversely polarized $\Lambda$ in the back-to-back production of  $\Lambda$ and a light hadron in $e^+ e^-$ collisions, $e^+ e^- \to \Lambda^\uparrow + h + X$~\cite{Guan:2018ckx}, from which the TMD PFFs have been extracted in~\cite{Callos:2020qtu,DAlesio:2020wjq}. 

We propose to study the transversely polarized $\Lambda$ production inside a jet in unpolarized proton-proton and lepton-proton collisions, $p+e/p\to (\text{jet}\,\Lambda^\uparrow)+ X$. We measure the longitudinal momentum fraction $z_\Lambda$ of the jet carried by the $\Lambda$ particle, transverse momentum ${\bm j}_\perp$ with respect to the jet direction, and the transverse spin ${\bm S}_{h\perp}$ of the $\Lambda$ particle. As emphasized already in section~\ref{sub:general}, the transverse momentum ${\bm j}_\perp$ and the transverse spin ${\bm S}_{h\perp}$ of the $\Lambda$ particle correlate with each other, generating a $\sin(\hat{\phi}_h - \hat{\phi}_{S_h})$ azimuthal dependence as in 
\bea
\frac{d\sigma}{dp_{JT} d\eta_{J} dz_\Lambda d^2{\bm j}_\perp}=&F_{UU,U}+|{\bm S}_{h\perp}| \sin(\hat{\phi}_h-\hat{\phi}_{S_h})F^{\sin(\hat{\phi}_h-\hat{\phi}_{S_h})}_{UU,T}+\cdots\,,
\eea
where ``$\cdots$'' represent other terms in eq.~\eqref{eq:gen_str}. We define the so-called $\Lambda$ transverse polarization observable inside the jet as follows
\bea
P_\Lambda = \frac{F^{\sin(\hat{\phi}_h-\hat{\phi}_{S_h})}_{UU,T}}{F_{UU,U}}\,.
\eea
Here the structure functions in the numerator and denominator can be written as the following factorized form~\footnote{once again, expressions for $ep$ follow simply.}
\bea
F_{UU,U}&=\sum_{a,b,c}\int_{x_a^{\text{min}}}^1\frac{dx_a}{x_a}f_a(x_a,\mu)\int_{x_b^{\text{min}}}^1\frac{dx_b}{x_b}f_b(x_b,\mu)\int_{z_c^{\text{min}}}^1\frac{dz_c}{z_c^2}H_{ab}^c \,\mathcal{D}_1^{\Lambda/c}(z_c,z_\Lambda,{\bm j}_\perp,p_{JT}R,\mu)\,,
\\
\label{eq:PFFTMDFJF}
F_{UU,T}^{\sin(\hat{\phi}_h-\hat{\phi}_{S_h})}
&=\sum_{a,b,c}\int_{x_a^{\text{min}}}^1\frac{dx_a}{x_a}f_a(x_a,\mu)\int_{x_b^{\text{min}}}^1\frac{dx_b}{x_b}f_b(x_b,\mu)\int_{z_c^{\text{min}}}^1\frac{dz_c}{z_c^2}H_{ab}^c \,\mathcal{D}_{1T}^{\perp\, \Lambda/c}(z_c,z_\Lambda,{\bm j}_\perp,p_{JT}R,\mu)\,.
\eea
Note that the expression for the unpolarized structure $F_{UU,U}$ has been given previously in~\cite{Kang:2017glf}, following which one can derive the formalism for the spin-dependent structure function~$F_{UU,T}^{\sin(\hat{\phi}_h-\hat{\phi}_{S_h})}$. In other words, they are sensitive to the unpolarized TMDJFFs $\mathcal{D}_1^{\Lambda/c}$ and the spin-dependent TMDJFFs $\mathcal{D}_{1T}^{\perp \Lambda/c}$, respectively. Their RG evolution equations and connections to the standard TMDFFs are given in section~\ref{sub:connection}. In particular, $\mathcal{D}_{1T}^{\perp \Lambda/c}$ is related to the TMD PFFs $D_{1T}^{\perp \Lambda/c}(z_h, {\bm j}_\perp; \mu_J)$ in eq.~\eqref{eq:PFFs}. 

In order to provide some estimate for $\Lambda$ transverse polarization inside the jet, we implement the following model for the TMD PFFs. Using the so-called $b_*$ prescription \cite{Collins:1984kg}, we combine TMD evolution with the recent gaussian fit of the Belle data to parametrize $D_{1T}^{\perp \Lambda/c}$ at the jet scale $\mu_J\sim p_{JT}R$ as
\bea
D_{T}^{\perp \Lambda/c}(z_\Lambda,{\bm j}_\perp;\mu_J)=\frac{1}{z^2_\Lambda}\left(\frac{1}{2z_\Lambda}\right)\int_0^{\infty}\frac{b^2\text{d}b}{2\pi}J_1\left(\frac{j_\perp b}{z_\Lambda}\right)F_c(z_\Lambda,\mu_{b_*})\ e^{-S_{\text{pert}}^i(b_*,\ \mu_J)-S_{\text{NP}}^i(b,\ \mu_J)},
\eea
where $S_{\text{pert}}^i$ are the usual perturbative Sudakov factors~\cite{Collins:2011zzd}, and  $F_c(z_\Lambda,\mu_{b_*})$ is fitted from the recent Belle data and has the following functional form 
\bea
\label{eq:Fc}
F_c(z_\Lambda,\mu_b^*) \equiv \mathcal{N}_c(z_\Lambda)\,D_1^{\Lambda/c}(z_\Lambda,\mu_b^*)\,,
\eea
where $D_{1}^{\Lambda/c}(z_\Lambda, \mu_b^*)$ are the unpolarized collinear $c\to \Lambda$ fragmentation functions, and the parametrization of $\mathcal{N}_c(z_\Lambda)$ for different quark flavors can be found in~\cite{Callos:2020qtu}. Note that at the moment, there is no existing extraction for the gluon TMD PFF $D_{1T}^{\perp \Lambda/g}$ and we thus do not include it in the numerical study below. However, we do include the unpolarized gluon TMDFF $D_{1}^{\Lambda/g}$ in the calculation of $F_{UU,U}$. The scale $\mu_{b_*} = 2e^{-\gamma_E}/{b_*}$ with $b_* =b/\sqrt{1+b^2/b^2_{\text{max}}}$. We then choose to parametrize the non-perturbative Sudakov factor for quark TMD PFFs as~\cite{Kang:2017glf,Kang:2015msa,Su:2014wpa}
\bea
S^q_{\text{NP}}(b,\mu_J)&=\frac{g_2}{2}\ln\left(\frac{b}{b_*}\right)\ln\left(\frac{p_{JT} R}{Q_0}\right)+\frac{\langle M_D^2\rangle}{4z_\Lambda^2}b^2\,,
\eea
with $Q_0^2 = 2.4~\text{GeV}^2,\ b_{\text{max}} = 1.5~\text{GeV}^{-1},\ g_2 = 0.84$, and $\langle M_D^2\rangle=0.118$~GeV$^2$~\cite{Callos:2020qtu}.

\begin{figure}
\centering
\begin{subfigure}{.33\textwidth}
  \centering
  \includegraphics[width=.99\linewidth]{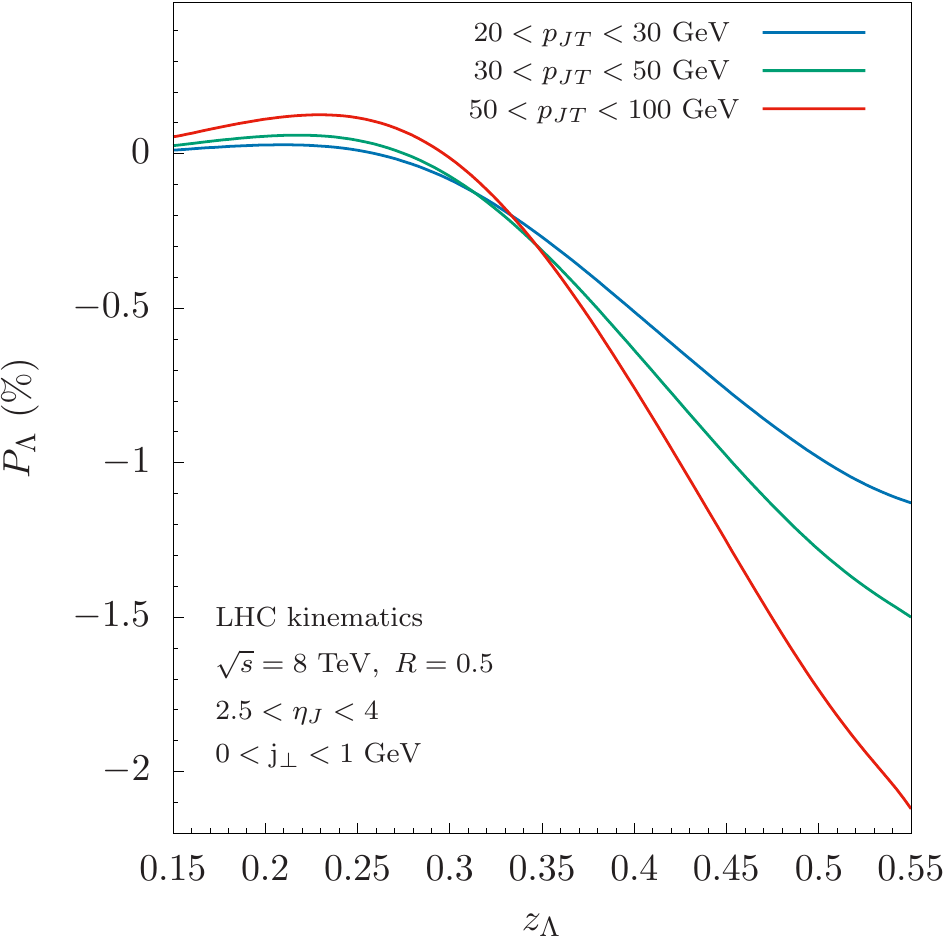} 
  \caption{}
  \label{fig:Plhcb}
\end{subfigure}
\begin{subfigure}{.33\textwidth}
  \centering
  \includegraphics[width=.99\linewidth]{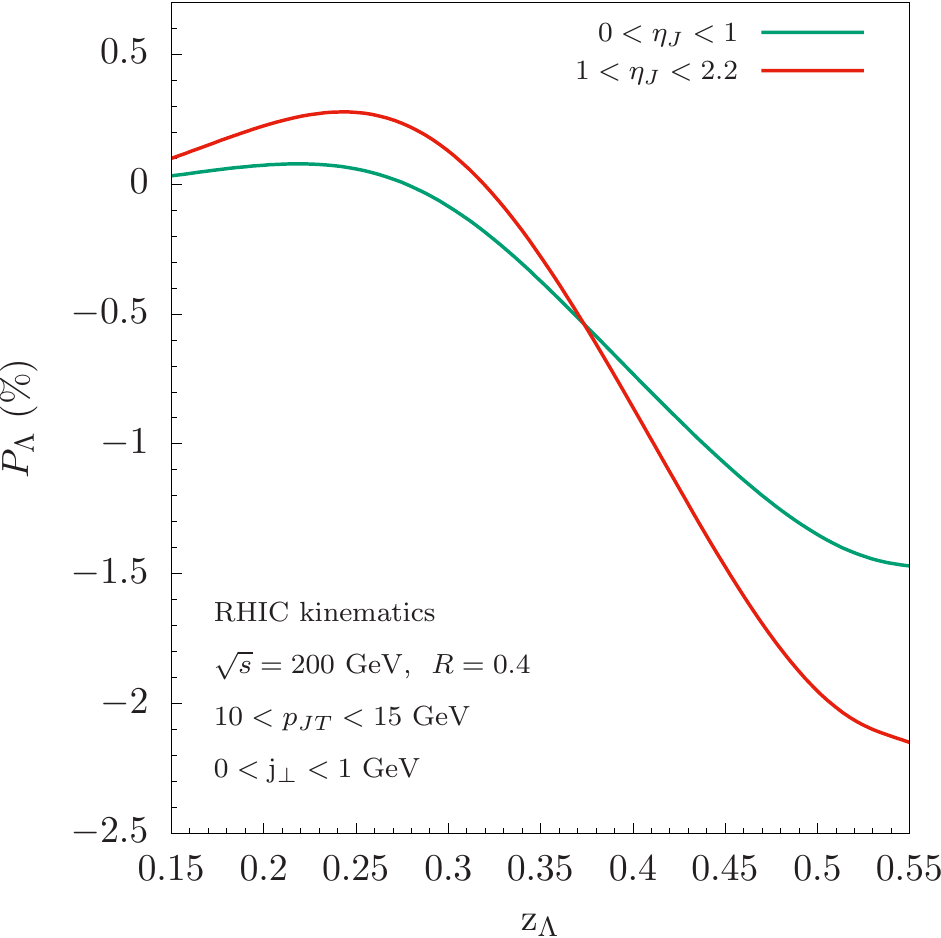} 
  \caption{}
    \label{fig:Ppp}
\end{subfigure}%
\begin{subfigure}{.33\textwidth}
  \centering
  \includegraphics[width=.99\linewidth]{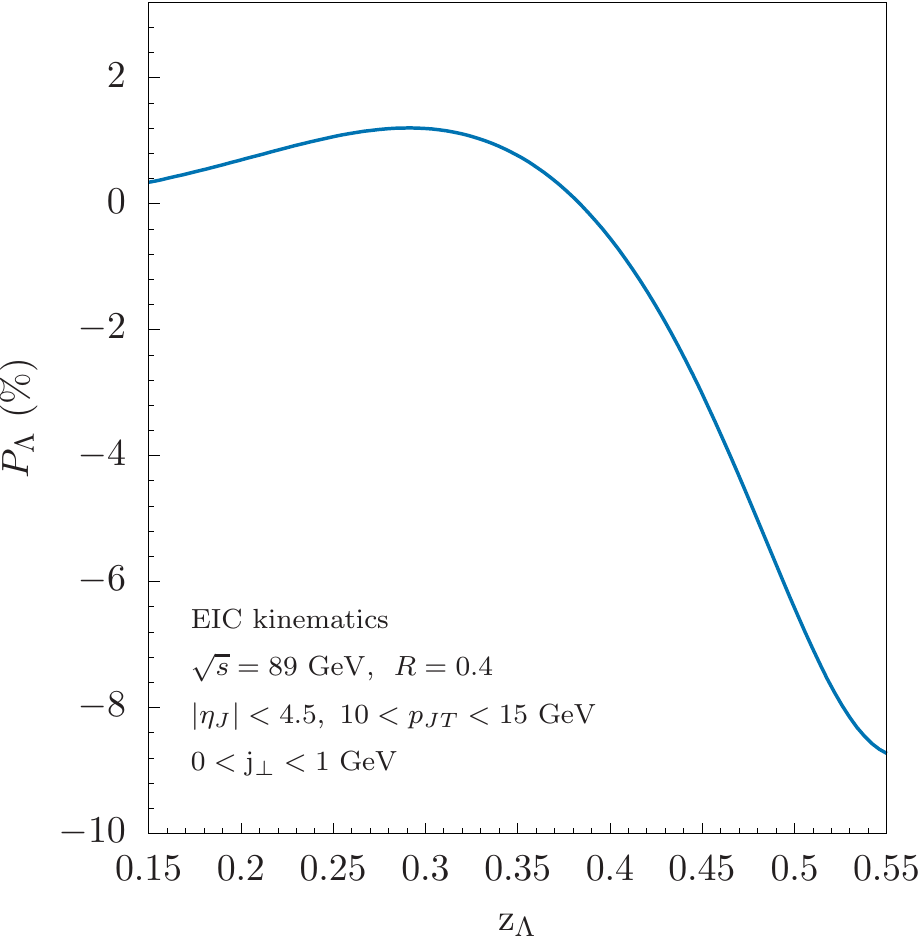} 
  \caption{}
  \label{fig:Pep}
\end{subfigure}
\caption{Predictions for the asymmetry $P_\Lambda$ for the LHC (left), RHIC (middle) and EIC (right) kinematics.}
\end{figure}

We now present our predictions for $\Lambda$ polarization $P_\Lambda$ at the LHC, RHIC and the future EIC. We use anti-$k_T$ jets with $R = 0.5$ for the LHC kinematics and $R = 0.4$ for the RHIC and EIC kinematics. We again use LO NNPDF \cite{ball2013parton} as the PDF sets and AKK08 \cite{Albino:2008fy} parametrization of the $\Lambda$ fragmentations to be consistent with the extraction of TMD PFFs we are using~\cite{Callos:2020qtu}. For the LHC, we implement the kinematics used on the recent LHCb measurements for the distribution of charged hadrons in $Z$-tagged jets~\cite{Aaij:2019ctd} in proton-proton collisions at the CM energy $\sqrt{s}=8~$TeV in the forward rapidity regions $2.5 < \eta_J < 4$. Fig.~\ref{fig:Plhcb} shows $z_\Lambda$ distribution of $P_\Lambda$ with $j_\perp$ integrated from $0 < j_\perp < 1~ \text{GeV}$ for three different ranges of jet transverse momenta: $20~<p_{JT} <~30$ GeV, $30~<p_{JT} <~50$ GeV and $50~<p_{JT} <~100$ GeV. The asymmetry $P_\Lambda$ starts out positive and becomes negative around $z_\Lambda \sim 0.3$. This behavior can be explained by the fact that PFF for the $u$ quark is positive from the extraction~\cite{Callos:2020qtu} and dominates in the small $z_\Lambda$ region. When $z_\Lambda \gtrsim 0.3$, negative PFF of the $d$ quark begins dominating. As the quark jet fraction increases with $p_{JT}$, we observe that asymmetry $P_\Lambda$ is enhanced as jet transverse momenta are increased. At the RHIC kinematics, we choose the transverse momentum of the jets to be $10~<p_{JT} <~15$ GeV and study two different ranges of rapidity, $0 < \eta_J < 1$ and $1 < \eta_J < 2.2$, where the latter may be available once a forward detector upgrade is made available at sPHENIX~\cite{Adare:2015kwa}. In fig.~\ref{fig:Ppp}, we present our results differential in $z_\Lambda$ again with $j_\perp$ integrated over $0 < j_\perp < 1~ \text{GeV}$. Just as $P_\Lambda$ at the LHC, $u$ quark PFF dominates when $z_\Lambda \lesssim 0.3$, but is taken over by negative $d$ quark PFF at the larger values of $z_\Lambda$. We also note that valence contributions are enhanced when we look at more forward rapidity region of $1 < \eta_J < 2.2$. Since valence quarks, especially $u$ and $d$, have the largest PFFs, size of $P_\Lambda$ is enhanced for the more forward rapidity region. Nevertheless, the $\Lambda$ polarization $P_{\Lambda}$ in proton-proton collisions at both LHC and RHIC is about $2\%$ level in magnitude, which is similar in size to the recent ATLAS measurement for the transverse polarization of single inclusive $\Lambda$ production in $pp\to \Lambda^\uparrow+X$~\cite{ATLAS:2014ona}. 

Finally, we present our prediction for $P_\Lambda$ as a function of $z_\Lambda$ for the future EIC at a CM energy of $\sqrt{s}=89~$GeV in fig.~\ref{fig:Pep}. We choose the transverse momentum and rapidity of the jets to be within the intervals $10~<p_{JT} <~15$ GeV and $|\eta_J| < 2$, respectively, and we again integrate $j_\perp$ over $0 < j_\perp < 1~ \text{GeV}$. While the result remains qualitatively similar to the ones for the LHC and RHIC kinematics, $u$ quark PFF dominates over a larger region of $z_\Lambda$ and $P_\Lambda$ remains positive until $z_\Lambda \lesssim 0.4$. Since quark PFFs dominate in the lepton-proton collisions, the polarization is larger in size with $P_\Lambda\sim 10\%$ at $z_\Lambda\sim 0.5$. Thus this would be a feasible jet substructure observable at the future EIC.


\section{Conclusion}
\label{sec:conc}
In this paper, we provide the general theoretical framework to study the distribution of hadrons inside the jet by taking full advantage of the polarization effects. We define the polarized jet fragmentation functions, in which both the parton that initiates the jet and the hadron inside the jet can have the general polarization. For single inclusive jet production in both proton-proton and lepton-proton collisions, we write down the general structure functions for hadron distribution inside the jet with different characteristic modulation in the azimuthal angles. We then use the developed frameworks to study both longitudinally polarized collinear $\Lambda$ production and transversely polarized transverse momentum dependent $\Lambda$ production. By carrying out phenomenological studies in the context of both $pp$ collider like LHC and RHIC and $ep$ collider like the EIC, we demonstrate sizeable asymmetry can be measured at different collider facilities. Our results especially show the importance of studying these observables at the future Electron-Ion Collider designed to shed light on hadron structures. We expect our work will open new and exciting opportunities in the direction of studying spin-dependent hadron structures using hadrons inside jets. 

\section*{Acknowledgments}
We thank B.~Page and J.~Zhang for useful discussions related to the longitudinally polarized $\Lambda$ measurement made by the STAR collaboration, and thank D.~Callos and J.~Terry for helpful discussions. Z.K. and F.Z. are supported by the National Science Foundation under Grant No.~PHY-1720486. K.L. is supported by the National Science Foundation under Grants No.~PHY-1316617 and No.~PHY-1620628.

\bibliographystyle{h-physrev5}
\bibliography{bibliography}

\begin{thebibliography}{10}

\bibitem{Procura:2009vm}
M.~Procura and I.~W. Stewart,
\newblock Phys. Rev. {\bf D81}, 074009 (2010), arXiv:0911.4980,
\newblock [Erratum: Phys. Rev.D83,039902(2011)].

\bibitem{Jain:2011xz}
A.~Jain, M.~Procura, and W.~J. Waalewijn,
\newblock JHEP {\bf 05}, 035 (2011), arXiv:1101.4953.

\bibitem{Jain:2011iu}
A.~Jain, M.~Procura, and W.~J. Waalewijn,
\newblock JHEP {\bf 04}, 132 (2012), arXiv:1110.0839.

\bibitem{Chien:2015ctp}
Y.-T. Chien, Z.-B. Kang, F.~Ringer, I.~Vitev, and H.~Xing,
\newblock JHEP {\bf 05}, 125 (2016), arXiv:1512.06851.

\bibitem{Arleo:2013tya}
F.~Arleo, M.~Fontannaz, J.-P. Guillet, and C.~L. Nguyen,
\newblock JHEP {\bf 04}, 147 (2014), arXiv:1311.7356.

\bibitem{Kaufmann:2015hma}
T.~Kaufmann, A.~Mukherjee, and W.~Vogelsang,
\newblock Phys. Rev. {\bf D92}, 054015 (2015), arXiv:1506.01415.

\bibitem{Kang:2016ehg}
Z.-B. Kang, F.~Ringer, and I.~Vitev,
\newblock JHEP {\bf 11}, 155 (2016), arXiv:1606.07063.

\bibitem{Dai:2016hzf}
L.~Dai, C.~Kim, and A.~K. Leibovich,
\newblock Phys. Rev. {\bf D94}, 114023 (2016), arXiv:1606.07411.

\bibitem{Kang:2019ahe}
Z.-B. Kang, K.~Lee, J.~Terry, and H.~Xing,
\newblock Phys. Lett. {\bf B798}, 134978 (2019), arXiv:1906.07187.

\bibitem{Kang:2017glf}
Z.-B. Kang, X.~Liu, F.~Ringer, and H.~Xing,
\newblock JHEP {\bf 11}, 068 (2017), arXiv:1705.08443.

\bibitem{Bacchetta:2000jk}
A.~Bacchetta and P.~J. Mulders,
\newblock Phys. Rev. {\bf D62}, 114004 (2000), arXiv:hep-ph/0007120.

\bibitem{Mulders:2000sh}
P.~J. Mulders and J.~Rodrigues,
\newblock Phys. Rev. {\bf D63}, 094021 (2001), arXiv:hep-ph/0009343.

\bibitem{Metz:2016swz}
A.~Metz and A.~Vossen,
\newblock Prog. Part. Nucl. Phys. {\bf 91}, 136 (2016), arXiv:1607.02521.

\bibitem{Neill:2016vbi}
D.~Neill, I.~Scimemi, and W.~J. Waalewijn,
\newblock JHEP {\bf 04}, 020 (2017), arXiv:1612.04817.

\bibitem{Neill:2018wtk}
D.~Neill, A.~Papaefstathiou, W.~J. Waalewijn, and L.~Zoppi,
\newblock JHEP {\bf 01}, 067 (2019), arXiv:1810.12915.

\bibitem{Makris:2017arq}
Y.~Makris, D.~Neill, and V.~Vaidya,
\newblock JHEP {\bf 07}, 167 (2018), arXiv:1712.07653.

\bibitem{Makris:2018npl}
Y.~Makris and V.~Vaidya,
\newblock JHEP {\bf 10}, 019 (2018), arXiv:1807.09805.

\bibitem{Gutierrez-Reyes:2019msa}
D.~Gutierrez-Reyes, Y.~Makris, V.~Vaidya, I.~Scimemi, and L.~Zoppi,
\newblock JHEP {\bf 08}, 161 (2019), arXiv:1907.05896.

\bibitem{Neill:2020bwv}
D.~Neill and F.~Ringer,
\newblock (2020), arXiv:2003.02275.

\bibitem{Kaufmann:2019ksh}
T.~Kaufmann, X.~Liu, A.~Mukherjee, F.~Ringer, and W.~Vogelsang,
\newblock JHEP {\bf 02}, 040 (2020), arXiv:1910.11746.

\bibitem{Bain:2016clc}
R.~Bain {\em et~al.},
\newblock JHEP {\bf 06}, 121 (2016), arXiv:1603.06981.

\bibitem{Anderle:2017cgl}
D.~P. Anderle, T.~Kaufmann, M.~Stratmann, F.~Ringer, and I.~Vitev,
\newblock Phys. Rev. D {\bf 96}, 034028 (2017), arXiv:1706.09857.

\bibitem{Kang:2017yde}
Z.-B. Kang, J.-W. Qiu, F.~Ringer, H.~Xing, and H.~Zhang,
\newblock Phys. Rev. Lett. {\bf 119}, 032001 (2017), arXiv:1702.03287.

\bibitem{Bain:2017wvk}
R.~Bain, L.~Dai, A.~Leibovich, Y.~Makris, and T.~Mehen,
\newblock Phys. Rev. Lett. {\bf 119}, 032002 (2017), arXiv:1702.05525.

\bibitem{Kaufmann:2016nux}
T.~Kaufmann, A.~Mukherjee, and W.~Vogelsang,
\newblock Phys. Rev. {\bf D93}, 114021 (2016), arXiv:1604.07175.

\bibitem{Aaboud:2019oac}
ATLAS, M.~Aaboud {\em et~al.},
\newblock Phys. Rev. Lett. {\bf 123}, 042001 (2019), arXiv:1902.10007.

\bibitem{Aad:2011td}
ATLAS, G.~Aad {\em et~al.},
\newblock Phys. Rev. {\bf D85}, 052005 (2012), arXiv:1112.4432.

\bibitem{Aad:2011sc}
ATLAS, G.~Aad {\em et~al.},
\newblock Eur. Phys. J. {\bf C71}, 1795 (2011), arXiv:1109.5816.

\bibitem{Chatrchyan:2012gw}
CMS, S.~Chatrchyan {\em et~al.},
\newblock JHEP {\bf 10}, 087 (2012), arXiv:1205.5872.

\bibitem{Chatrchyan:2014ava}
CMS, S.~Chatrchyan {\em et~al.},
\newblock Phys. Rev. {\bf C90}, 024908 (2014), arXiv:1406.0932.

\bibitem{Aad:2014wha}
ATLAS, G.~Aad {\em et~al.},
\newblock Phys. Lett. {\bf B739}, 320 (2014), arXiv:1406.2979.

\bibitem{Aaboud:2017tke}
ATLAS, M.~Aaboud {\em et~al.},
\newblock Nucl. Phys. {\bf A978}, 65 (2018), arXiv:1706.02859.

\bibitem{Aaij:2017fak}
LHCb, R.~Aaij {\em et~al.},
\newblock Phys. Rev. Lett. {\bf 118}, 192001 (2017), arXiv:1701.05116.

\bibitem{Acharya:2019zup}
ALICE, S.~Acharya {\em et~al.},
\newblock JHEP {\bf 08}, 133 (2019), arXiv:1905.02510.

\bibitem{Aaij:2019ctd}
LHCb, R.~Aaij {\em et~al.},
\newblock Phys. Rev. Lett. {\bf 123}, 232001 (2019), arXiv:1904.08878.

\bibitem{Acharya:2018eat}
ALICE, S.~Acharya {\em et~al.},
\newblock Phys. Rev. D {\bf 99}, 012016 (2019), arXiv:1809.03232.

\bibitem{Aad:2019onw}
ATLAS, G.~Aad {\em et~al.},
\newblock Phys. Rev. D {\bf 100}, 052011 (2019), arXiv:1906.09254.

\bibitem{Sirunyan:2019vlp}
CMS, A.~M. Sirunyan {\em et~al.},
\newblock Phys. Lett. B {\bf 804}, 135409 (2020), arXiv:1910.01686.

\bibitem{Adamczyk:2017wld}
STAR, L.~Adamczyk {\em et~al.},
\newblock Phys. Rev. D {\bf 97}, 032004 (2018), arXiv:1708.07080.

\bibitem{Adamczyk:2017ynk}
STAR, L.~Adamczyk {\em et~al.},
\newblock Phys. Lett. {\bf B780}, 332 (2018), arXiv:1710.10215.

\bibitem{Accardi:2012qut}
A.~Accardi {\em et~al.},
\newblock Eur. Phys. J. A {\bf 52}, 268 (2016), arXiv:1212.1701.

\bibitem{Aschenauer:2017jsk}
E.~Aschenauer {\em et~al.},
\newblock Rept. Prog. Phys. {\bf 82}, 024301 (2019), arXiv:1708.01527.

\bibitem{Liu:2018trl}
X.~Liu, F.~Ringer, W.~Vogelsang, and F.~Yuan,
\newblock Phys. Rev. Lett. {\bf 122}, 192003 (2019), arXiv:1812.08077.

\bibitem{Arratia:2019vju}
M.~Arratia, Y.~Song, F.~Ringer, and B.~Jacak,
\newblock (2019), arXiv:1912.05931.

\bibitem{Yuan:2007nd}
F.~Yuan,
\newblock Phys. Rev. Lett. {\bf 100}, 032003 (2008), arXiv:0709.3272.

\bibitem{DAlesio:2010sag}
U.~D'Alesio, F.~Murgia, and C.~Pisano,
\newblock Phys. Rev. D {\bf 83}, 034021 (2011), arXiv:1011.2692.

\bibitem{DAlesio:2011kkm}
U.~D'Alesio, L.~Gamberg, Z.-B. Kang, F.~Murgia, and C.~Pisano,
\newblock Phys. Lett. B {\bf 704}, 637 (2011), arXiv:1108.0827.

\bibitem{Kang:2017btw}
Z.-B. Kang, A.~Prokudin, F.~Ringer, and F.~Yuan,
\newblock Phys. Lett. B {\bf 774}, 635 (2017), arXiv:1707.00913.

\bibitem{DAlesio:2017bvu}
U.~D'Alesio, F.~Murgia, and C.~Pisano,
\newblock Phys. Lett. B {\bf 773}, 300 (2017), arXiv:1707.00914.

\bibitem{Bauer:2000ew}
C.~W. Bauer, S.~Fleming, and M.~E. Luke,
\newblock Phys. Rev. {\bf D63}, 014006 (2000), arXiv:hep-ph/0005275.

\bibitem{Bauer:2000yr}
C.~W. Bauer, S.~Fleming, D.~Pirjol, and I.~W. Stewart,
\newblock Phys. Rev. {\bf D63}, 114020 (2001), arXiv:hep-ph/0011336.

\bibitem{Bauer:2001ct}
C.~W. Bauer and I.~W. Stewart,
\newblock Phys. Lett. {\bf B516}, 134 (2001), arXiv:hep-ph/0107001.

\bibitem{Bauer:2001yt}
C.~W. Bauer, D.~Pirjol, and I.~W. Stewart,
\newblock Phys. Rev. {\bf D65}, 054022 (2002), arXiv:hep-ph/0109045.

\bibitem{Meissner:2007rx}
S.~Meissner, A.~Metz, and K.~Goeke,
\newblock Phys. Rev. {\bf D76}, 034002 (2007), arXiv:hep-ph/0703176.

\bibitem{Diehl:2001pm}
M.~Diehl,
\newblock Eur. Phys. J. {\bf C19}, 485 (2001), arXiv:hep-ph/0101335.

\bibitem{Diehl:2003ny}
M.~Diehl,
\newblock Phys. Rept. {\bf 388}, 41 (2003), arXiv:hep-ph/0307382.

\bibitem{Goeke:2005hb}
K.~Goeke, A.~Metz, and M.~Schlegel,
\newblock Phys. Lett. {\bf B618}, 90 (2005), arXiv:hep-ph/0504130.

\bibitem{Mulders:1995dh}
P.~J. Mulders and R.~D. Tangerman,
\newblock Nucl. Phys. {\bf B461}, 197 (1996), arXiv:hep-ph/9510301,
\newblock [Erratum: Nucl. Phys.B484,538(1997)].

\bibitem{Altarelli:1977zs}
G.~Altarelli and G.~Parisi,
\newblock Nucl. Phys. {\bf B126}, 298 (1977).

\bibitem{Stratmann:1996hn}
M.~Stratmann and W.~Vogelsang,
\newblock Nucl. Phys. {\bf B496}, 41 (1997), arXiv:hep-ph/9612250.

\bibitem{KangLeeZhao:inprogress}
Z.-B. Kang, K.~Lee, and F.~Zhao,
\newblock in progress  (2020).

\bibitem{Cacciari:2008gp}
M.~Cacciari, G.~P. Salam, and G.~Soyez,
\newblock JHEP {\bf 04}, 063 (2008), arXiv:0802.1189.

\bibitem{Buffing:2018ggv}
M.~G. Buffing, Z.-B. Kang, K.~Lee, and X.~Liu,
\newblock (2018), arXiv:1812.07549.

\bibitem{Aidala:2012mv}
C.~A. Aidala, S.~D. Bass, D.~Hasch, and G.~K. Mallot,
\newblock Rev. Mod. Phys. {\bf 85}, 655 (2013), arXiv:1209.2803.

\bibitem{deFlorian:1997zj}
D.~de~Florian, M.~Stratmann, and W.~Vogelsang,
\newblock Phys. Rev. {\bf D57}, 5811 (1998), arXiv:hep-ph/9711387.

\bibitem{Burkardt:1993zh}
M.~Burkardt and R.~L. Jaffe,
\newblock Phys. Rev. Lett. {\bf 70}, 2537 (1993), arXiv:hep-ph/9302232.

\bibitem{Buskulic:1996vb}
ALEPH, D.~Buskulic {\em et~al.},
\newblock Phys. Lett. {\bf B374}, 319 (1996).

\bibitem{deFlorian:1998ba}
D.~de~Florian, M.~Stratmann, and W.~Vogelsang,
\newblock Phys. Rev. Lett. {\bf 81}, 530 (1998), arXiv:hep-ph/9802432.

\bibitem{Xu:2006my}
STAR, Q.-h. Xu,
\newblock AIP Conf. Proc. {\bf 915}, 428 (2007), arXiv:hep-ex/0612035.

\bibitem{Abelev:2009xg}
STAR, B.~I. Abelev {\em et~al.},
\newblock Phys. Rev. {\bf D80}, 111102 (2009), arXiv:0910.1428.

\bibitem{Adam:2018kzl}
STAR, J.~Adam {\em et~al.},
\newblock Phys. Rev. {\bf D98}, 112009 (2018), arXiv:1808.07634.

\bibitem{ball2013unbiased}
R.~D. Ball {\em et~al.},
\newblock Nuclear Physics B {\bf 874}, 36 (2013).

\bibitem{ball2013parton}
R.~D. Ball {\em et~al.},
\newblock Nuclear Physics B {\bf 867}, 244 (2013).

\bibitem{Guan:2018ckx}
Belle, Y.~Guan {\em et~al.},
\newblock Phys. Rev. Lett. {\bf 122}, 042001 (2019), arXiv:1808.05000.

\bibitem{Callos:2020qtu}
D.~Callos, Z.-B. Kang, and J.~Terry,
\newblock (2020), arXiv:2003.04828.

\bibitem{DAlesio:2020wjq}
U.~D'Alesio, F.~Murgia, and M.~Zaccheddu,
\newblock (2020), arXiv:2003.01128.

\bibitem{Collins:1984kg}
J.~C. Collins, D.~E. Soper, and G.~F. Sterman,
\newblock Nucl. Phys. {\bf B250}, 199 (1985).

\bibitem{Collins:2011zzd}

\newblock J.~Collins{\em {Foundations of perturbative QCD}} Vol.~32 (Cambridge
  University Press, 2013).

\bibitem{Kang:2015msa}
Z.-B. Kang, A.~Prokudin, P.~Sun, and F.~Yuan,
\newblock Phys. Rev. {\bf D93}, 014009 (2016), arXiv:1505.05589.

\bibitem{Su:2014wpa}
P.~Sun, J.~Isaacson, C.~P. Yuan, and F.~Yuan,
\newblock Int. J. Mod. Phys. {\bf A33}, 1841006 (2018), arXiv:1406.3073.

\bibitem{Albino:2008fy}
S.~Albino, B.~A. Kniehl, and G.~Kramer,
\newblock Nucl. Phys. {\bf B803}, 42 (2008), arXiv:0803.2768.

\bibitem{Adare:2015kwa}
PHENIX, A.~Adare {\em et~al.},
\newblock (2015), arXiv:1501.06197.

\bibitem{ATLAS:2014ona}
ATLAS, G.~Aad {\em et~al.},
\newblock Phys. Rev. D {\bf 91}, 032004 (2015), arXiv:1412.1692.

\end{thebibliography}

\end{document}